\documentclass{emulateapj}
\usepackage{graphicx}
\usepackage{natbib}
\slugcomment{To appear in ApJ., XX.}
\shorttitle{Time-Varying \ion{K}{1} in SN\,2014J}
\shortauthors{Graham et al.}

\begin{document}
\title{Time-Varying Potassium in High-Resolution Spectra of the Type Ia Supernova 2014J}
\author{
M.~L. Graham\altaffilmark{1},
S. Valenti\altaffilmark{2,3},
B.~J. Fulton\altaffilmark{4},
L.~M. Weiss\altaffilmark{1},
K.~J. Shen\altaffilmark{1},
P.~L. Kelly\altaffilmark{1},
W. Zheng\altaffilmark{1},
A.~V. Filippenko\altaffilmark{1},
G.~W. Marcy\altaffilmark{1},
D.~A. Howell\altaffilmark{2,3},
J. Burt\altaffilmark{5}, and
E.~J. Rivera\altaffilmark{5}
}

\altaffiltext{1}{Department of Astronomy, University of California, Berkeley, CA 94720-3411, USA}
\altaffiltext{2}{Las Cumbres Observatory Global Telescope Network, Goleta CA 93117, USA}
\altaffiltext{3}{Physics Department, University of California, Santa Barbara CA 93106, USA}
\altaffiltext{4}{Institute for Astronomy, University of Hawaii, 2680 Woodlawn Dr, Honolulu, HI 96822, USA}
\altaffiltext{5}{UCO/Lick Observatory, Department of Astronomy and Astrophysics, University of California, Santa Cruz, CA 95064, USA}

\begin{abstract}

We present a time series of the highest resolution spectra yet published for the nearby Type Ia supernova (SN) 2014J in M82. They were obtained at 11 epochs over 33 days around peak brightness with the Levy Spectrograph (resolution $R\approx$ 110,000) on the 2.4\,m Automated Planet Finder telescope at Lick Observatory. We identify multiple \ion{Na}{1}~D and \ion{K}{1} absorption features, as well as absorption by \ion{Ca}{1} H\&K and several of the more common diffuse interstellar bands (DIBs). We see no evolution in any component of \ion{Na}{1}~D, \ion{Ca}{1}, or in the DIBs, but do establish the dissipation/weakening of the two most blueshifted components of \ion{K}{1}. We present several potential physical explanations, finding the most plausible to be photoionization of circumstellar material, and discuss the implications of our results with respect to the progenitor scenario of SN\,2014J.
\end{abstract}


\section{Introduction} \label{sec:intro}

Type Ia supernovae (SNe\,Ia), the thermonuclear explosions of carbon-oxygen white dwarf stars, are powerful cosmological standardizable candles, but their progenitor scenarios and their explosion mechanisms are not yet well understood (e.g., Howell 2011). The common progenitor options include the double-degenerate (DD) scenario of two white dwarf stars, and the single-degenerate (SD) scenario of one white dwarf with either a red giant or main sequence companion. Constraints on the nature of dark energy from SNe\,Ia are currently limited by the systematic uncertainty in the calibration of SNe\,Ia. A physical understanding of SNe\,Ia would help address these systematic uncertainties, in particular the relative contributions of intrinsic SN\,Ia color and dust absorption along the line of sight, both from the interstellar medium (ISM) and, potentially, circumstellar material (CSM).

It is unlikely that a white dwarf would accrete all of the mass lost by its companion, and the remainder may enshroud the system as CSM. The amount, composition, and velocity of this material is dependent on the progenitor scenario. From an observer's perspective, this outflowing material would appear blueshifted and may be close enough to interact with the SN\,Ia ejecta, causing evolution in the narrow spectral features (e.g., of H, Ca, or Na). The first identified example of this was SN\,Ia 2006X, for which a complex of blueshifted \ion{Na}{1}~D lines was seen to evolve between $-2$, $+14$, and $+61$ days relative to maximum brightness \citep{Patat2007}. The highest velocity components --- the material most recently released and closest to the SN --- were the first to show evolution. Two additional cases of time-variable Na were presented by Blondin et al. (2009) and Simon et al. (2009), respectively for high- and low-extinction SNe\,Ia, both concluding the presence of CSM. Another recent example is the nearby SN\,Ia PTF\,11kx, for which a time series of high-resolution spectra showing clear evolution in the blueshifted lines of Ca, Fe, He, and H painted a picture of the SN\,Ia progenitor star exploding into multiple shells of CSM \citep{Dilday2012}. 

In a compilation of high-resolution ($R \equiv \lambda/\delta \lambda \approx$ 30,000--50,000) spectra of 35 SNe\,Ia, Sternberg et al. (2011) show that narrow \ion{Na}{1}~D absorption lines are preferentially blueshifted, and that this is inconsistent with a random distribution of line-of-sight clouds; rather, it is better explained by the outflow of CSM from the progenitor system. Furthermore, Phillips et al. (2013) find that the visual extinction, $A_V$, is better correlated with the equivalent width (EW) of the diffuse interstellar band (DIB) at 5780\,\AA\ than with the EW of \ion{Na}{1}~D, indicating that not all of the \ion{Na}{1}~D absorption is caused by the ISM. Most recently, Sternberg et al. (2014) compile the largest sample yet of high-resolution spectra and find that $\sim18$\% exhibit time-variable Na, indicating the presence of CSM.

Together, these studies suggest that CSM may be more common in SNe\,Ia than previously thought, which may explain why the total-to-selective extinction ($R_V$)  derived from SN\,Ia light curves is sometimes significantly lower than common Milky Way sight-lines, especially for SNe\,Ia with high $A_V$ (e.g., Wang 2005). These low $R_V$, high $A_V$ SNe\,Ia exhibit high photospheric velocities more often than ``normal" SNe\,Ia, and occur more often in the central regions of larger host galaxies, associating them with a younger, more metal-rich stellar population \citep{Wang2013}. The latter suggests that they may have a different progenitor scenario, perhaps one more suited to the creation of CSM. However, this is a very tenuous connection to make, and recent work by Mandel et al. (2014) shows that the correlation between SN\,Ia color (i.e., $R_V$) and photospheric velocity is more likely to be caused by intrinsic color differences than by dust reddening. 

SN\,Ia spectra contain mostly broad features, upon which the relatively narrower absorption lines from the ISM are clearly seen. For this reason, bright SNe are very useful as probes of the extragalactic ISM \citep{Rich1987,DOdorico1989,Steidel1990,Sollerman2005,Patat2010,Welty2014,Ritchey2014}. For example, spectra of SN\,1987A in the Large Magellanic Cloud (LMC) marked the first detection of Li, Ca, and K in an external galaxy's ISM, revealed a different \ion{Na}{1} to \ion{Ca}{2} ratio than for the Milky Way, and provided a large collection of DIB observations \citep{Vladilo1987a,Vladilo1987b}. Additionally, evolution in the EW of \ion{Na}{1}~D absorption components could be used to constrain the sizes of clouds in the ISM as the SN photosphere expands behind them, as detailed by Patat et al. (2010).

SN\,Ia 2014J was discovered in the dusty irregular galaxy M82 in mid-January 2014. With a discovery magnitude of $R=10.5$ and a distance of just $\sim3.5$\,Mpc, SN\,2014J was accessible to a wider variety of telescopes and instruments than most SNe. Initial observations found that SN\,2014J exhibited high-velocity spectral features ($v \lesssim$ 16,000\,km\,s$^{-1}$) and significant extinction ($A_V=2.5\pm1.3$\,mag; Goobar et al. 2014). This makes SN\,2014J a useful case study for potential CSM and as a probe of the ISM of M82. In light of this, here we present a time series of high-resolution spectra obtained with the new Levy Spectrograph on the Automated Planet Finder at Lick Observatory \citep{Vogt2014}. 

The paper is organized as follows. In \S \ref{sec:2014J}, we begin with a review of what has been published so far about SN\,2014J. We present and analyze the new observations in \S~\ref{sec:obs}, including multiple resolved velocity components of the \ion{Na}{1}~D and \ion{K}{1} lines, the \ion{Ca}{2} H\&K lines, and multiple DIBs. Section~\ref{sec:disc} discusses the implications of our observations for SNe\,Ia and summarizes our conclusions.

\section{The Type Ia Supernova 2014J} \label{sec:2014J}

SN\,2014J was discovered (Fossey et al. 2014) in images taken on 21.8 January 2014 (UT dates are used throughout this paper), and classified as a Type Ia SN $\sim1$ week before peak brightness with a spectrum taken on 22.3 January 2014 (Cao et al. 2014). Early observations revealed SN\,2014J to be highly reddened and extinguished, with deep \ion{Na}{1}~D absorption, and a member of the high-velocity subgroup based on its \ion{Si}{2} $\lambda$6355 absorption line (up to 20,000\,km\,s$^{-1}$ reported; Fossey et al. 2014). The coordinates of SN\,2014J are $\alpha = 9^{\rm h}55^{\rm m}42.137^{\rm s}$, $\delta = +69^\circ40'25.40''$ (J2000.0; Kelly et al. 2014), located in the nearby galaxy M82 at distance $D = 3.52 \pm 0.02$\,Mpc (based on the tip of the red giant branch; Jacobs et al. 2009) and having a recession velocity of $v_{\rm rec}=203$\,km\,s$^{-1}$ (heliocentric; de Vaucouleurs et al. 1991). 

Within a week of discovery the first papers about this remarkable SN\,Ia were released. Zheng et al. (2014) present prediscovery photometry of SN\,2014J, showing that its brightness rise is best fit with a varying power law (whereas most early-time SN light curves are assumed to be $\propto t^2$), and constraining the time of first light to 14.75 January 2014. Their paper also presents the first optical spectrum of SN\,2014J, demonstrating its Type Ia classification. Goobar et al. (2014) present early-time data showing that SN\,2014J is spectroscopically normal but highly reddened by host-galaxy dust, and that the absorption lines of \ion{Na}{1}~D, \ion{Ca}{2} H\&K, and the DIB at 5780\,\AA\ do not exhibit any evolution in EW. They also find that the EW of the DIB at 5780\,\AA\ is $0.48 \pm 0.1$\,\AA, which via the relation established by Phillips et al. (2013)\footnote{quoted in our Equation 1 in \S 3.4} indicates an extinction of $A_{\rm host}=2.5\pm1.3$\,mag.

Amanullah et al. (2014) present {\it Hubble Space Telescope (HST)} multi-band photometry of SN\,2014J, and show that it reached a peak $B$-band magnitude on 1 February 2014. With their ultraviolet, optical, and infrared data they confirm a total-to-selective extinction ratio of $R_V = 1.4\pm0.1$, and rule out the typical Galactic value of $R_V \geq 3.1$. A low value of $R_V$ could be caused by smaller dust grains along the line of sight to SN\,2014J. This would agree with the results of Kawabata et al. (2014), whose optical spectropolarimetry shows that the dust grains in M82 are likely smaller than those in the Milky Way (as has been noted for other extinguished SNe\,Ia). Amanullah et al. (2014) also find that a power-law extinction, which is expected in the scenario where a dusty circumstellar environment causes multiple scatterings of the light (e.g., Wang 2005), agrees well with their observations. This conclusion of circumstellar dust is supported by the work of Foley et al. (2014), who combine photometry, {\it HST} ultraviolet to near-infrared spectra, and a time series of high-resolution spectra to show that the reddening and extinction observed for SN\,2014J stem from a $\sim 50/50$ combination of typical dust and scattering in the circumstellar medium. They find no evidence for evolution in the components of \ion{Na}{1}~D, \ion{K}{1}, or the DIBs in their spectroscopic time series. However, Patat et al. (2014) show that the continuum polarization in SN\,2014J is aligned with the local spiral structure of M82, and that the line-of-sight dust is most likely interstellar small grains, not circumstellar material.

Limits on the density of circumstellar material from X-ray and radio nondetections are presented by Margutti et al. (2014) and P\'{e}rez-Torres et al. (2014), respectively. In both cases, the CSM expected from most SD scenarios is ruled out. In the DD scenario where a white dwarf binary is surrounded by a uniform CSM, its density is limited to $\lesssim 1.0$\,cm$^{-3}$ within radius $R = 1.6 \times 10^{17}$\,cm. They also limit the mass-loss rate from a putative companion to $<10^{-9}$\,M$_{\odot}$\,yr$^{-1}$ (assuming the mass-loss rate is constant over time). For comparison, radio limits for SN\,2006X constrain the mass-loss rate to $<10^{-8}$\,M$_{\odot}$\,yr$^{-1}$, yet the evolving \ion{Na}{1}~D clearly indicates that CSM for the progenitor system was still detected \citep{Patat2007}. A recent analysis of {\it Spitzer Space Telescope} mid-infrared observations of SN\,2014J similarly rules out the presence of CSM within $\sim10^{17}$\,cm (Johansson et al. 2014).

Six epochs of high-resolution ($R \approx$ 30,000) optical spectra bracketing the time of peak brightness were obtained and examined in tandem papers by Welty et al. (2014) and Ritchey et al. (2014). These works catalog and analyze the complex of absorption features along the line of sight to SN\,2014J. They note that the bulk of the \ion{H}{1} emission (from radio data) along the line of sight to SN\,2014J is at a velocity of $\sim -70$\,km\,s$^{-1}$ with respect to the rest frame of M82, and optical spectra presented by Westmoquette et al. (2009) shows that the approaching side of the disk of M82 is southwest of its center. This means that SN\,2014J is on the approaching side of the disk. Welty et al. and Ritchey et al. model the absorption features as multiple components with a velocity resolution of $\sim 9.5$\,km\,s$^{-1}$, and find that compositional diversity between dust clouds in M82 can be broadly categorized in two groups: molecular gas that may be significantly shielded or experiencing a weaker radiation field, and atomic gas with either less shielding or a stronger ambient radiation field. The molecular gas has velocities from $-70$ to $-157$\,km\,s$^{-1}$ with respect to the rest frame of M82 (consistent with the bulk of \ion{H}{1}), while the atomic gas is $>-70$\,km\,s$^{-1}$. Welty et al. and Ritchey et al. find no evolution in any absorption feature over their time series of spectra.

Constraints on the progenitor scenario are provided by studies of pre-SN archival images of M82. Kelly et al. (2014) present archival {\it HST} images and Keck AO images to constrain the magnitude of a potential progenitor binary companion star, and are able to rule out a bright red giant or a cool He star companion. Nielsen et al. (2014) use archival Chandra data to show that the SD scenario including a unobscured supersoft X-ray source progenitor is not favored.

The emerging picture is of SN\,2014J as a high-velocity SN\,Ia, extinguished and reddened by a diverse collection of intervening material along the line of sight. Pre-SN observations, and upper limits from nondetections of SN\,2014J in radio and X-ray data, mostly rule out the predicted direct signatures of CSM from the SD progenitor scenarios. However, contributions from both ISM and CSM are inferred from the low value of $R_V$ derived from SN\,2014J observations, which could indicate either smaller dust grains in M82 or a dusty CSM causing multiple scatterings of the light. Evolution in the spectral absorption features that would indicate an interaction between the SN ejecta and the CSM, and thus be direct evidence of CSM physically associated with the progenitor system, have not yet been found.

\section{Analysis of the APF Spectra of SN\,2014J} \label{sec:obs}

We observed SN\,2014J on 11 epochs over 33 days between 22 January 2014 and 24 February 2014 with the 2.4\,m Automated Planet Finder (APF) telescope at Lick Observatory. The APF hosts the Levy Spectrograph, a high-resolution optical echelle spectrograph with $R$(5500\,\AA) $\approx$ 110,000 for a slit width of 1\arcsec\ \citep{Vogt2014}, which is a  velocity resolution of $\sim3$\,m\,s$^{-1}$. Although this is remarkably high resolution for observations of a supernova, our data do not set the record --- Pettini (1988) present $R \approx$ 600,000 spectra over \ion{Na}{1}~D for SN\,1987A from the Anglo-Australian Telescope. The observation dates, phase of SN\,2014J, and exposure times for each spectrum are listed in Table \ref{tab:spec}. In this work we use the date of $B$-band peak brightness reported by Foley et al. (2014), 2 February 2014.

\begin{deluxetable}{lccc}[b]
\tablecolumns{5}
\tablecaption{SN\,2014J APF Spectroscopy \label{tab:spec}}
\tablehead{
\colhead{Observation}                 & 
\colhead{SN\,Ia Phase}         & 
\colhead{Exposure}  \\
\colhead{UT Date}                 & 
\colhead{(days)}         & 
\colhead{Time (s)} 
} 
\startdata
2014-01-22 & $-11$ & $4 \times 1200$ \\
2014-01-23 & $-10$ & $8 \times 600$ \\
2014-01-25 & $-8$ & $3 \times 600$ \\
2014-01-26 & $-7$ & $3 \times 600$ \\
2014-01-27 & $-6$ & $3 \times 600$ \\
2014-02-01 & $-1$ & $3 \times 600$ \\
2014-02-05 & 3 & $3 \times 600$ \\
2014-02-15 & 13 & $3 \times 600$ \\
2014-02-17 & 15 & $3 \times 1200$ \\
2014-02-20 & 18 & $3 \times 1200$ \\
2014-02-24 & 22 & $3 \times 1200$
\enddata
\tablenotetext{}{}
\end{deluxetable}

To the spectral orders containing the Na, Ca, and K lines we apply a wavelength calibration and correct for barycentric velocity before converting to the rest frame of M82 using a recession velocity of $v_{\rm{rec}} = 203$\,km\,s$^{-1}$. The features of a SN\,Ia spectrum are too broad for measuring the peculiar velocity of SN\,2014J within M82, and so instead we work in the rest frame of M82. 

For the orders containing the identified DIBs, we have not found that it is necessary to do a precise wavelength calibration because we are mainly interested in their EW (the precise calibration is very time consuming as it must be done manually using night-sky lines, and we avoid it when possible). The flux in each spectral order is highly blazed, and in lieu of deblazing the spectra we present continuum-normalized fluxes, sufficient for our analysis. The spectral orders containing the Na and K lines have contributions from atmospheric absorption lines of $\rm H_2O$ or $\rm O_2$, which we fit and remove using a similar method to that presented for SN\,2011fe by Patat et al. (2013).

In the following sections we present and analyze the absorption lines for Na, Ca, K, and a few DIBs. There are several atomic and molecular interstellar lines commonly seen and analyzed in other publications with high-resolution spectroscopy that our observations do not cover. In particular, our spectra do not extend sufficiently blue to see \ion{Ti}{2} $\lambda$3383.76 or the molecular lines of CN, for which the band head is at 3883.4\,\AA. The rotational bands of CH$^+$ are at micron wavelengths, also not covered by our data. The second line of the potassium doublet \ion{K}{1} $\lambda\lambda$7664.90, 7698.96 unfortunately falls into a gap between orders. We see no evidence of the DIBs at 4428\,\AA\ and 6379\,\AA, which have been reported in other publications. The \ion{Li}{1} $\lambda$6707 line detected by Ritchey et al. (2014) falls at the very edge of an order, where the flux is low, and we do not see this feature. Finally, we do not find any evidence of the hydrogen Balmer lines.

\subsection{Sodium: \ion{Na}{1}~D} \label{ss:na}

\begin{figure*}[t]
\begin{center}
\includegraphics[trim=3cm 0.5cm 4cm 1cm,clip=true,width=18cm]{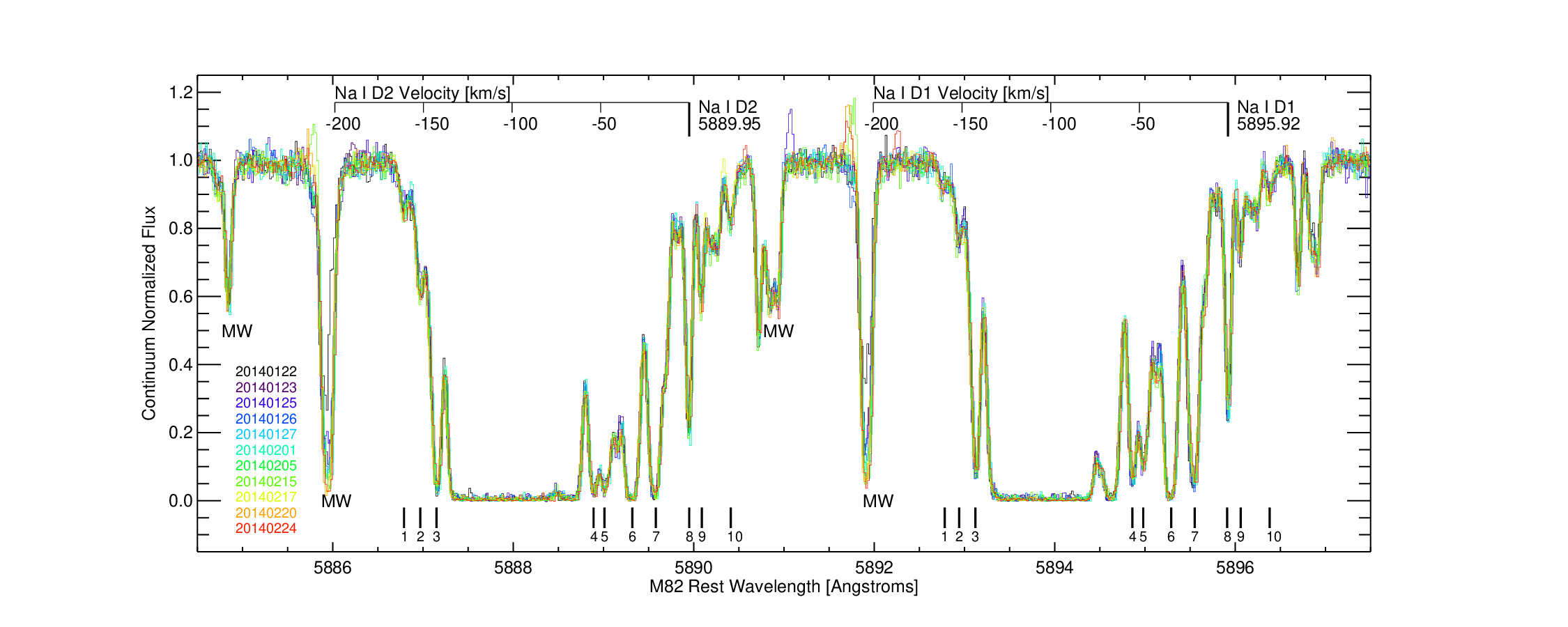}
\includegraphics[trim=3cm 0.5cm 4cm 1cm,clip=true,width=18cm]{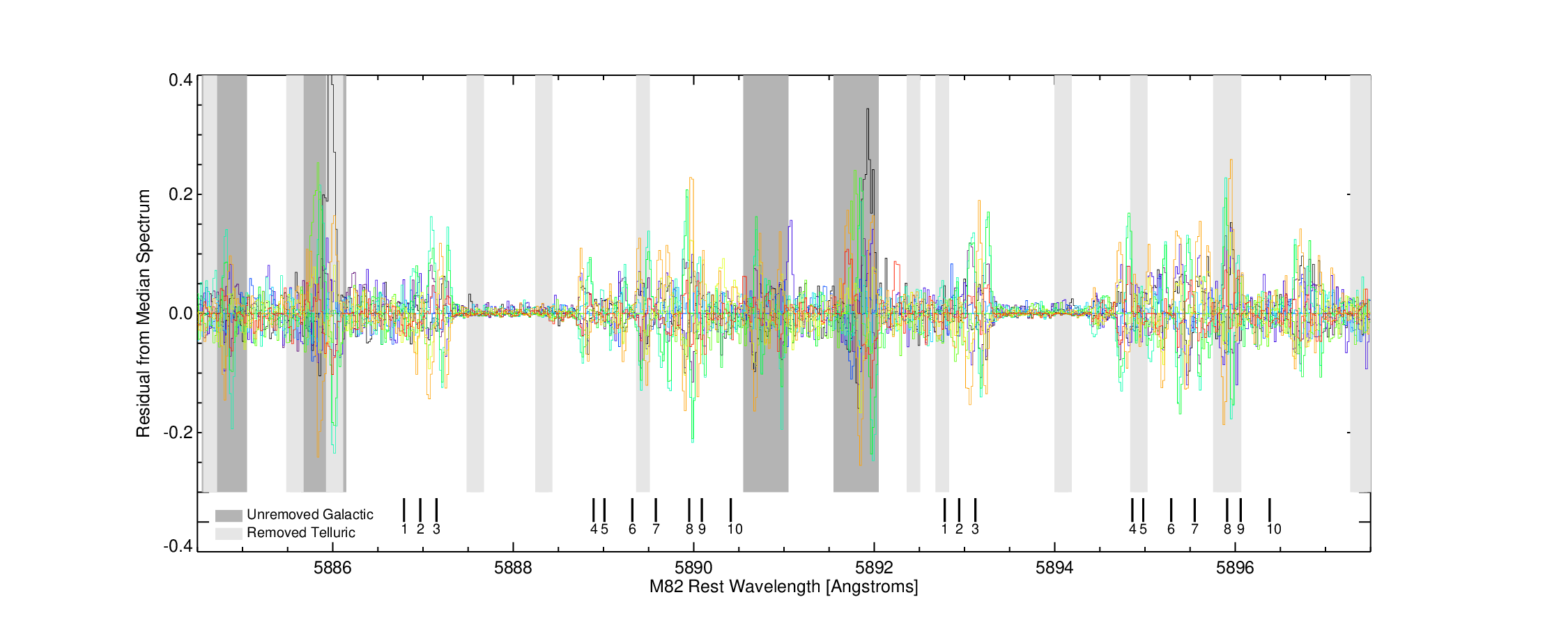}
\caption{{\it Top:} APF spectra of SN\,2014J in the region of the \ion{Na}{1}~D line for all dates (represented with colors, listed at left). The wavelengths have been calibrated and corrected for the redshift of M82 and the Earth's barycentric velocity. The fluxes have been continuum normalized, and the telluric features have been fit and removed. The Milky Way components of \ion{Na}{1}~D are labeled with ``MW.'' Short vertical bars at the top mark the wavelengths of \ion{Na}{1}~D in the rest frame of M82, with a scale bar showing component velocity. Short vertical bars along the bottom mark the 10 individual unsaturated components of \ion{Na}{1}~D listed in Table \ref{tab:NaDpair}. {\it Bottom:} Residuals from the median spectrum, which was created using all epochs. Light-gray background marks atmospheric H$_2$O (width 0.15\,\AA) and darker gray marks Galactic sodium (width 0.5\,\AA). Larger residuals from the median spectrum appear to correlate with nonsaturated sodium features. \label{fig:NaD_pano}}
\end{center}
\end{figure*}

In the top frame of Figure \ref{fig:NaD_pano} we show the time series of high-resolution APF spectra around the wavelengths of \ion{Na}{1}~D. As in Goobar et al. (2014), we identify two Galactic absorption components, labeled ``MW'' (Milky Way) in Figure \ref{fig:NaD_pano}: a deeper component at $v\approx 0$ and a shallower one at $v\approx -50$\,km\,s$^{-1}$ (in the observer's frame; $-203$ and $-253$\,km\,s$^{-1}$ in the M82 rest frame used in Figure \ref{fig:NaD_pano}). The component at 0\,km\,s$^{-1}$ appears to vary, but this represents the difficulty in subtracting the strong atmospheric sodium emission at 0\,km\,s$^{-1}$; scattering of light from low-pressure sodium lamps in cities near Lick Observatory is variable. Further evidence of this is visible in the noisy residuals at the edges of the line. The extreme saturation of the \ion{Na}{1}~D feature was expected from the severe host-galaxy reddening and extinction observed for SN\,2014J in low-resolution spectra. We identify 10 nonsaturated components, numbered in Figure \ref{fig:NaD_pano}, and we list their wavelengths and velocity relative to the rest frame of M82 in Table \ref{tab:NaDpair}. The components span a total velocity range of 183\,km\,s$^{-1}$. 

\begin{deluxetable}{ccc}[tb]
\tablecolumns{3}
\tablecaption{Velocity Components of \ion{Na}{1} D \label{tab:NaDpair}}
\tablehead{
\colhead{Line}                 & 
\colhead{Wavelengths}         & 
\colhead{Velocity\tablenotemark{a}} \\
\colhead{ID \#}                 & 
\colhead{(\AA)}         & 
\colhead{(km\,s$^{-1}$)}
} 
\startdata
1  & 5886.79, 5892.78 & $-160$ \\
2  & 5886.97, 5892.94 & $-152$ \\
3  & 5887.50, 5893.12 & $-143$ \\
4  & 5888.89, 5894.86 & $-54$ \\
5  & 5889.01, 5894.98 & $-48$ \\
6  & 5889.32, 5895.29 & $-32$ \\
7  & 5889.58, 5895.55 & $-19$ \\
8  & 5889.95, 5895.91 & 0 \\
9  & 5890.09, 5896.06 & +7 \\
10 & 5890.41, 5896.38 & +23
\enddata
\tablenotetext{a}{With respect to the rest frame of M82.}
\tablenotetext{}{}
\end{deluxetable}

All \ion{Na}{1}~D absorption components could be consistent with dust clouds along the line of sight in the ISM of M82. Optical spectroscopy with Gemini's Integral Field Unit (IFU) of M82 has shown that the approaching side of the disk is to the southwest of the center, which is the region in which SN\,2014J is located (e.g., Figure 18 of Westmoquette et al. 2009). The intervening ISM features are thus expected to be blueshifted. Further IFU data suggest that the dominant component of \ion{Na}{1}~D has a velocity in the rest frame of M82 of $v\approx-80$\,km\,s$^{-1}$ (based on an extrapolation of the data presented in Figure 4 of Westmoquette et al. 2013), which corresponds to the center of the saturated component in our spectra and the bulk of the \ion{H}{1} emission at radio wavelengths presented by Ritchey et al. (2014). In the case of SN\,2014J, it seems that the blueshifted \ion{Na}{1}~D absorption complex is consistent with expectations for the ISM of M82, but blueshifted \ion{Na}{1}~D with velocities $-50$ to $-200$\,km\,s$^{-1}$ has also been associated with SN\,Ia CSM, originating as material released from the progenitor system and/or swept up ISM (e.g., Sternberg et al. 2011; Maguire et al. 2013). Although commonly attributed to the SD system, Shen et al. (2013) have shown that blueshifted \ion{Na}{1}~D is also a natural byproduct of a He WD companion in the DD scenario. 

Is this \ion{Na}{1}~D line representative of the ISM of M82, the CSM of SN\,2014J, or both? In none of the IFU data presented by Westmoquette et al. (2013) is the \ion{Na}{1}~D line saturated like it is for SN\,2014J, which may indicate a ``special" sightline (i.e., contaminated by extra material such as CSM). Here we look for additional evidence that these components might be associated with the CSM of SN\,2014J. First, we note that the components nearest to the rest wavelength of M82, \#8 and \#9, have a full width at half-maximum intensity (FWHM) of 4--5 pixels. This is significantly narrower than two of the most unblended and unsaturated blueshifted lines, \#3 and \#7, that have a FWHM of 7--8 pixels. Owing to blending, it is difficult to determine a distribution of line widths, but it appears that the gas which creates the blueshifted absorption components is different from that near the rest frame of M82. This agrees with the interpretation of two broad types of ISM in M82 containing molecular and atomic gas, the former blueshifted with respect to the rest frame of M82 (Ritchey et al. 2014).

\begin{figure}[t]
\begin{center}
\epsscale{1.2}
\plotone{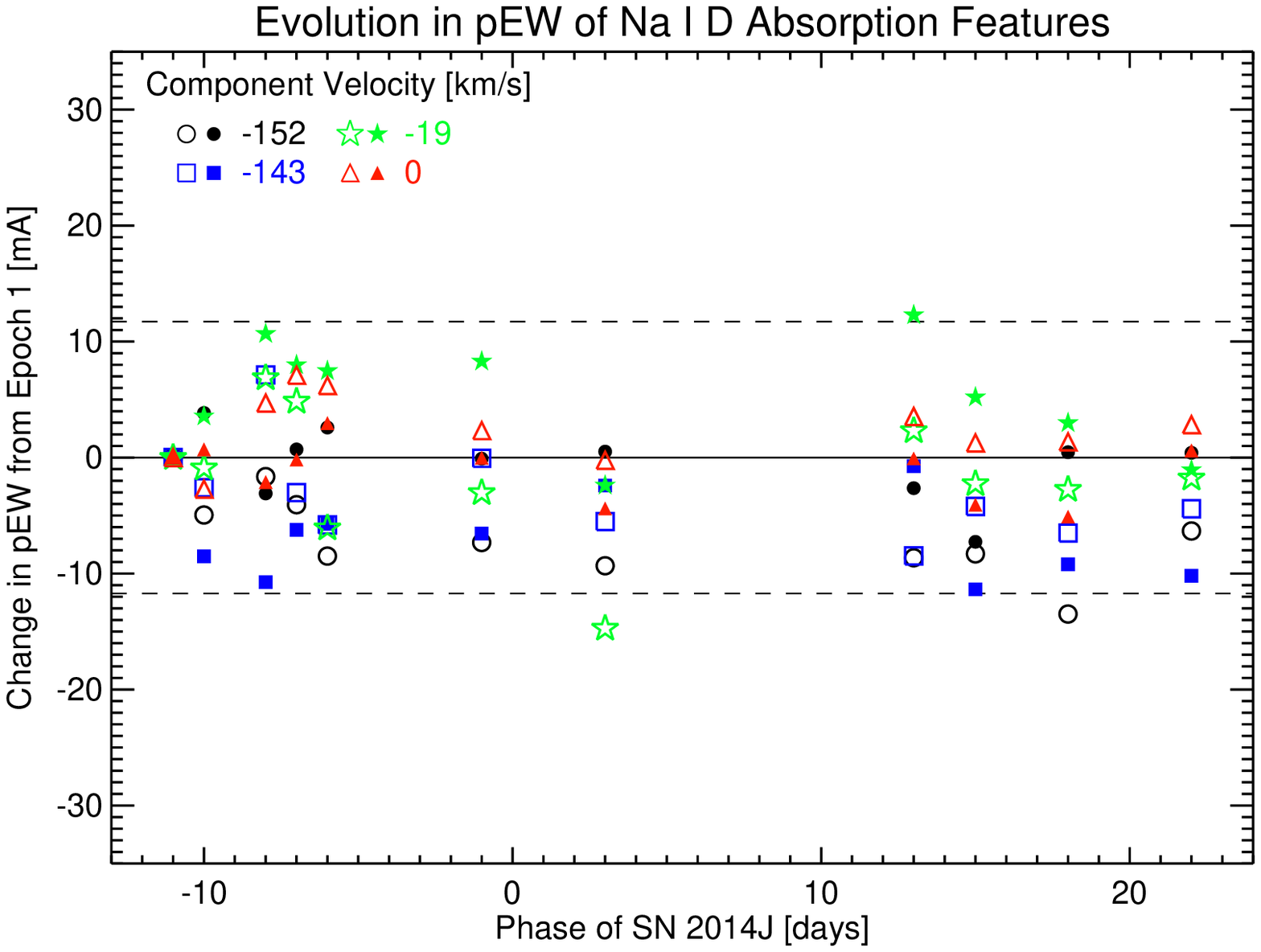}
\caption{Change in pseudo-equivalent width (pEW) between $-11$ and +22 days for four of the most unblended components of \ion{Na}{1}~D: line \#2 (black circles), 3 (blue squares), 7 (green stars), and 8 (red triangles; all are listed in Table \ref{tab:NaDpair}). Open and filled symbols represent \ion{Na}{1}~D1 and \ion{Na}{1}~D2, respectively. Horizontal bars mark the average uncertainty in pEW over all measurements. None of these lines shows significant or consistent evolution for both \ion{Na}{1}~D components in our time series of spectra. \label{fig:pairs}}
\end{center}
\end{figure}

Second, we look for evidence of evolution in the absorption features. Our final spectrum was obtained $\sim40$ days after explosion; ejecta material traveling at typical SN\,Ia speeds of 10,000\,km\,s$^{-1}$ would encompass a $3.5 \times 10^{15}$\,cm radius by then. If any of these components are within that distance, we should see changes to both \ion{Na}{1}~D lines. In the bottom panel of Figure \ref{fig:NaD_pano} we plot flux residuals from the median spectrum, which was created by taking the median flux over all epochs at each wavelength. We note that larger residual fluxes are seen at the wavelengths of nonsaturated sodium features --- but is this consistent with the uncertainty in our measurements, or representative of real physical evolution in the line-of-sight material? If the variation in absorption-line flux is caused by physical evolution in the line-of-sight material, we would expect both the \ion{Na}{1}~D1 and D2 components to vary in a consistent, monotonic fashion. In Figure \ref{fig:pairs} we present the pseudo-equivalent width (pEW\footnote{A linear pseudo-continuum is created by connecting the line edges, and then we calculate the pEW. We do this to compensate for contributions from neighboring, blended lines. This is appropriate for our purposes because these pEW values are only compared internally to look for relative changes over time, and not used as an absolute measurement of the EW.}) as a function of time for four of the most unblended \ion{Na}{1}~D lines,  \#2, 3, 7, and 8 in Table \ref{tab:NaDpair}. None of these show a consistent evolution in the pEW that would indicate a physical change related to the supernova. 

\subsection{Potassium: \ion{K}{1}} \label{ss:k}

\begin{figure}[t]
\begin{center}
\epsscale{2.4}
\plottwo{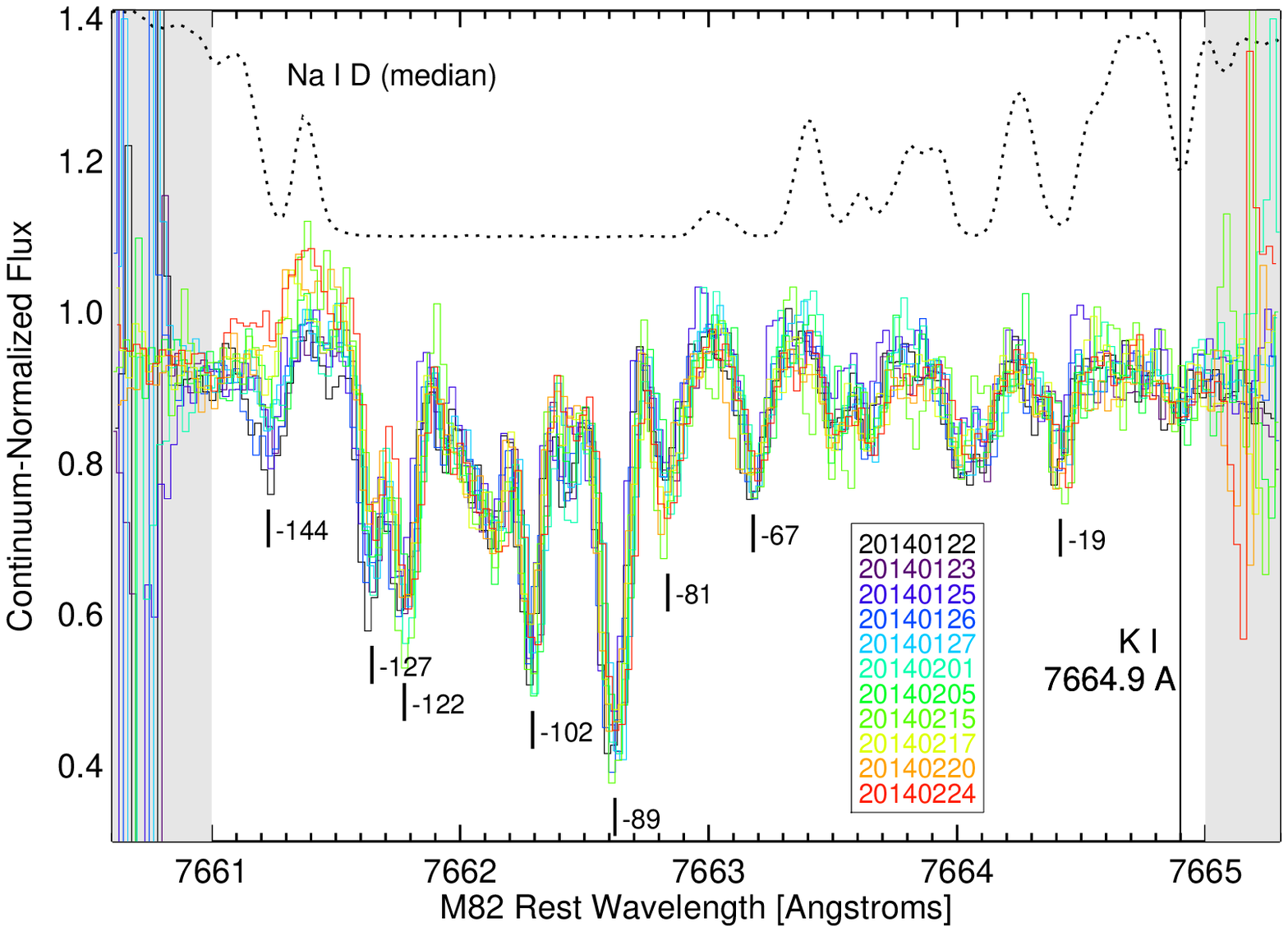}{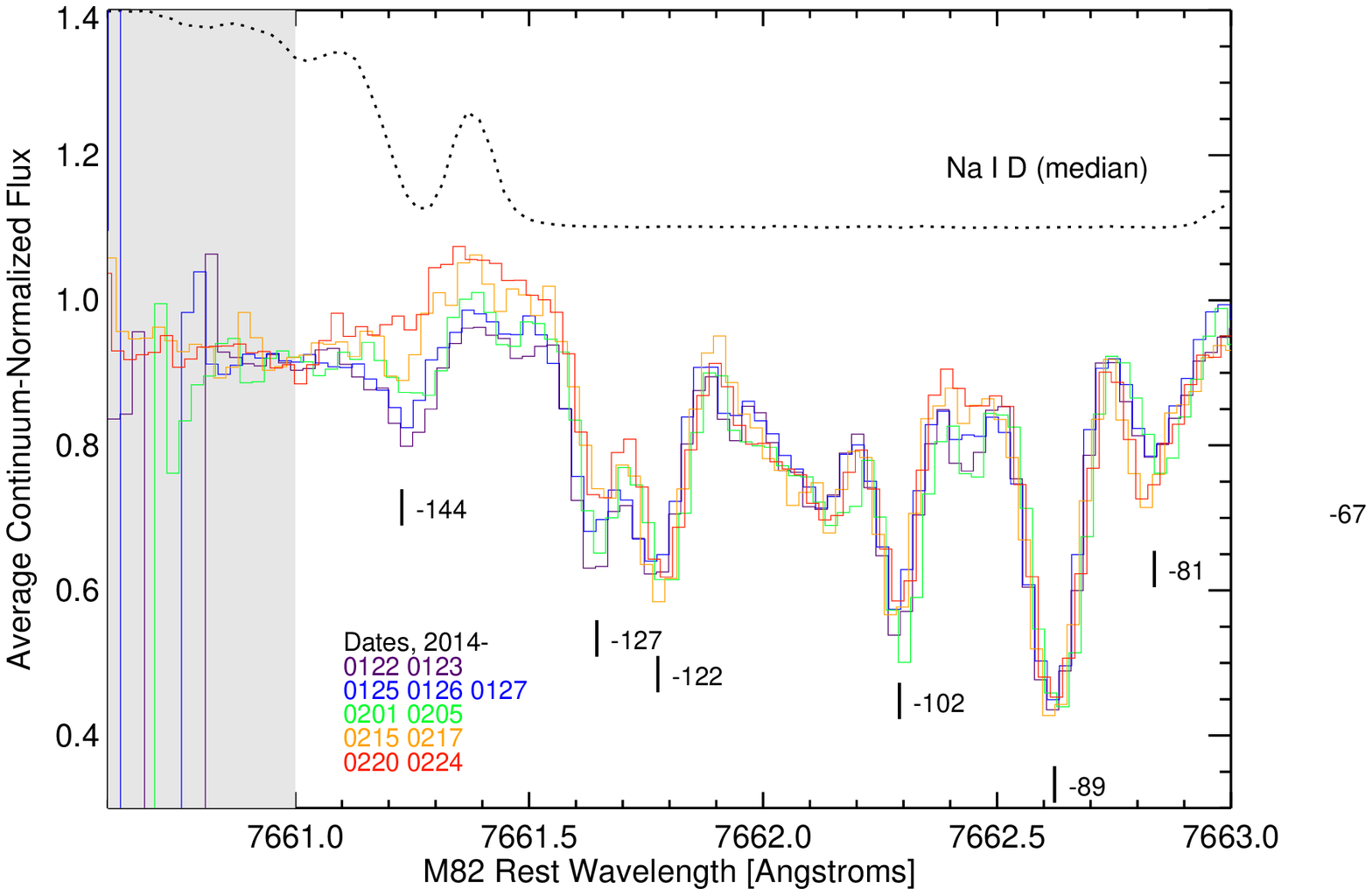}
\caption{The potassium line \ion{K}{1} $\lambda$7665. Vertical solid line marks the position of \ion{K}{1}. Telluric lines have been fit and subtracted, in some cases leaving noisy residuals (gray background regions). Lines unblended enough to individually identify are marked with short vertical lines and their line velocity in km\,s$^{-1}$. For comparison, the dotted spectrum along the top of the plot shows the \ion{Na}{1}~D1 absorption feature, matched to \ion{K}{1} in velocity space. In the top plot, the time series progresses from black through purple, blue, green, yellow, and orange to the red line. In this we see consistent evolution (shrinking) of the absorption lines at $v\approx-144$ and $-127$\,km\,s$^{-1}$. The dissipation of these lines is clearer in the bottom plot, where we have grouped the spectra by date into five epochs (colored lines) and zoomed in on the most blueshifted features to show the evolution in average continuum-normalized flux. \label{fig:K1}}
\end{center}
\end{figure}

The \ion{K}{1} doublet lines are at $\lambda = 7664.90$ and 7698.96\,\AA, and unfortunately the latter falls into a gap between spectral orders. In Figure \ref{fig:K1} we show the absorption complex created by \ion{K}{1} $\lambda$7664.90, for each epoch of our time series. The atmospheric lines have been fit and subtracted, leaving noisy residuals which we mark with a gray background. For comparison the \ion{Na}{1}~D feature from Figure \ref{fig:NaD_pano} is drawn across the top of the plot, velocity matched to \ion{K}{1}.

Three qualities immediately jump out of Figure \ref{fig:K1}. First, the saturated feature of \ion{Na}{1}~D is not saturated in \ion{K}{1}. The \ion{K}{1} complex clearly has multiple components with a velocity spacing of 10--20\,km\,s$^{-1}$ that extend up to $\sim -150$\,km\,s$^{-1}$, matching the \ion{Na}{1}~D very well. This velocity range also agrees with previous observations of CSM (e.g., $\sim100$\,km\,s$^{-1}$ for SN\,Ia PTF\,11kx; Dilday et al. 2012) and expectations from theoretical models (e.g., $\sim100$\,km\,s$^{-1}$; Shen et al. 2013). Second, the absorption line at the rest wavelength of M82 is relatively much less significant, almost nonexistent, for \ion{K}{1} compared to \ion{Na}{1}~D. The relative depths of the blueshifted components of \ion{K}{1} are similar to their corresponding velocity component in \ion{Na}{1}~D, suggesting a different composition between the material in the rest frame of M82 and the blueshifted material. This was also presented by Ritchey et al. (2014). Third, consistent and monotonic evolution appears in the most blueshifted features of \ion{K}{1}, in particular the $v \approx -144$ and $v \approx -127$\,km\,s$^{-1}$ components.

\begin{figure}[t]
\begin{center}
\includegraphics[trim=0cm 1cm 0.5cm 2.1cm,clip=true,width=8cm]{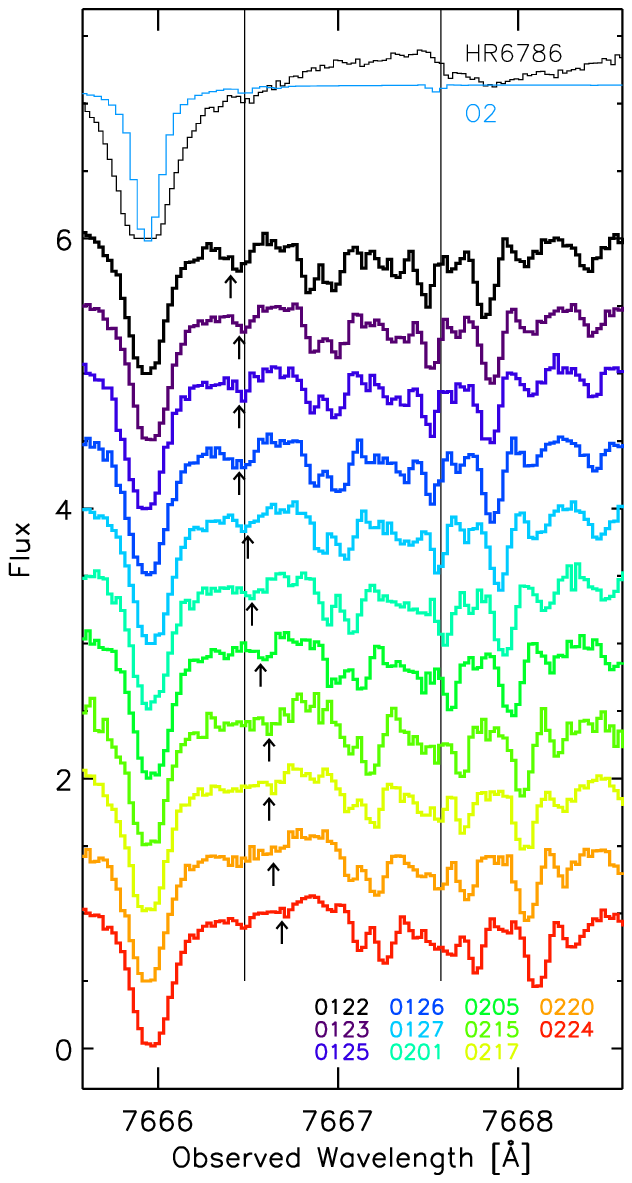}
\caption{A time series of the \ion{K}{1} line in raw data, as described in the text. Fluxes have been scaled and offset for clarity. Colors represent individual epochs, with the dates listed along the bottom (same colors as in Figure \ref{fig:K1}). Vertical lines mark the positions of two subtle telluric lines, and the upward arrow points to the expected position of the most blueshifted line ($v\approx-144$\,km\,s$^{-1}$ with respect to the rest frame of M82). Along the top we show a synthetic atmospheric O$_2$ spectrum in light blue, and an APF spectrum of the ISM standard star HR6786 in black. \label{fig:K1raw}}
\end{center}
\end{figure}

True evolution in the most blueshifted features of \ion{K}{1} would be a remarkable claim, and we must meticulously assess whether this observation is real, an artifact of the data-reduction process, or related to atmospheric phenomena. We have examined the two-dimensional images and find no nearby atmospheric emission lines that may introduce variation during sky subtraction (e.g., such as in the $v\approx0$\,km\,s$^{-1}$ Galactic component of \ion{Na}{1}~D in Figure \ref{fig:NaD_pano}). This region of the spectrum is noisier than that of \ion{Na}{1}~D, and the features of \ion{K}{1} are not as clearly static as those of \ion{Na}{1}~D, but none of the others exhibit such consistent and \textit{monotonic} evolution as the two most blueshifted features. In the one-dimensional spectra shown in Figure \ref{fig:K1}, the region between the two most blueshifted lines also varies, perhaps indicating a problem with the continuum normalization at \ion{K}{1}. These blueshifted lines are also close to large subtracted telluric features. For these reasons, in Figure \ref{fig:K1raw} we plot the time series of one-dimensional spectra prior to the telluric subtraction, continuum normalization, and barycentric velocity correction (i.e., sky subtraction only).

At the top of Figure \ref{fig:K1raw} we show, for comparison, an APF spectrum of ISM standard star HR6786 and a synthetic O$_2$ spectrum \citep{Patat2013}. Vertical lines mark the positions of two subtle telluric lines, and the upward arrow points to the expected position of the most blueshifted line ($v\approx-144$\,km\,s$^{-1}$ with respect to the rest frame of M82; the arrow moves right for successive epochs due to barycentric velocity). In the earliest phases, the relative contributions from the small telluric line and the feature at $v\approx-144$\,km\,s$^{-1}$ at the position of the arrow are unclear. By 5 February 2014 that feature is distinctly removed from the telluric, and by 17 February 2014 it appears to have dissipated almost entirely. Although the strengths of atmospheric lines can vary site to site and night to night, the telluric line at this position is expected to be much smaller than the \ion{K}{1} absorption feature. To assess potential night-to-night variations in that region, we examined the spectra of standard stars taken on the same night as our data, but they all appear very similar to HR6786. 

There are two possible observational interpretations for the most blueshifted ($v\approx-144$\,km\,s$^{-1}$) component of the \ion{K}{1} line. {\bf (1)} It does not exist, and our observations are instead a combination of an unexpectedly strong telluric feature in the first $\sim4$--5 epochs and noise in the remaining epochs. {\bf (2)} It has dissipated over time between 22 January 2014 and 24 February 2014, corresponding to phases $-11$ and +22 days with respect to peak brightness (or +8 and +41 days since ``first light" as defined by Zheng et al. 2014). Regarding option (1), we point out that there are two potential physical reasons why this component may exist in \ion{Na}{1}~D but not \ion{K}{1}: {\bf (1a)} the absorbing material is similar to that creating the $v\approx0$\,km\,s$^{-1}$ component, which does not appear to be significant in \ion{K}{1} either (Figure \ref{fig:K1}), or {\bf (1b)} the material is circumstellar and its potassium has experienced photoionization or collisional ionization within the first 8 days after explosion. However, we judge option (1) to be less likely, given the relative weakness of the interloping telluric line and strength of this component in \ion{Na}{1}~D.

We find that option (2), true dissipation of the feature, to be more likely --- especially given the partial dissipation of the second-most blueshifted ($v \approx -127$\,km\,s$^{-1}$) feature (compare with the $v \approx -122$\,km\,s$^{-1}$ feature), at a position where no telluric contamination is expected. Furthermore, it is important to note that we cannot rule out a small change of $\sim10$\,m\AA\ in the corresponding \ion{Na}{1}~D component (see the pEW evolution of the $v\approx-143$\,km\,s$^{-1}$ component of \ion{Na}{1}~D in Figure \ref{fig:pairs}). The initial pEW of the $v\approx-144$\,km\,s$^{-1}$ feature in \ion{K}{1} is $\sim30$\,m\AA, and so a physical process that affects $\lesssim1$ \ion{Na}{1} atom for every 3 \ion{K}{1} atoms may be plausible. Here we consider three potential physical explanations for our observations: photoionization of CSM by SN near-ultraviolet (NUV) radiation, collisional ionization of CSM by SN ejecta, and transverse motions perpendicular to our line of sight to SN\,2014J.

\textbf{Photoionization --} Could this evolution in \ion{K}{1} represent ionization by photons from SN\,2014J? The ionization energies for these two species are similar, $E$(\ion{Na}{1}) = 5.14\,eV and $E$(\ion{K}{1}) = 4.34\,eV \citep{NIST}, and the wavelengths of the ionizing photons are $\lambda$(\ion{Na}{1}) $\approx 2400$\,\AA\ and $\lambda$(\ion{K}{1}) $\approx 2850$\,\AA. In the case where material is ionized by NUV photons from SN\,2014J, it is important to note that the NUV spectra of SNe\,Ia can be quite steep. For example, at a phase of $-10$ days with respect to peak brightness, the NUV spectrum of SN\,Ia 2011fe exhibited a factor of $\sim12.5$ more flux at $\lambda\approx2850$\,\AA\ than at $\lambda\approx2400$\,\AA\footnote{Reduced {\it HST} NUV-optical spectra from Mazzali et al. (2014) are available on WISEREP (Yaron \& Gal-Yam 2012).}. Two caveats here are that (1) this difference decreases to a factor of 3.3 by the time of the light-curve peak, and (2) the presence of a high photospheric velocity in SN\,2014J may indicate that its NUV spectrum differs from that of SN\,2011fe \citep{Milne2013}. The intrinsic NUV behavior of SN\,2014J is difficult to derive because it depends on the true value of $R_V$ (e.g., Foley et al. 2014). 
Here we use the NUV spectra of SN\,2011fe from Mazzali et al. (2014) to examine whether it is feasible that \ion{K}{1} is ionized by photons from SN\,2014J, when none or only a small amount of \ion{Na}{1} is. The cumulative number of photoionizations that have occurred of species X by time $t^{\prime}$, $P_{\rm{X}}(t^{\prime})$, is given by

\begin{equation}
P_{\rm{X}}(t^{\prime}) \approx \int_{t_0}^{t^{\prime}} \alpha_{\rm{X}}\ N_{\rm{X}}(t)\ \frac{D^2}{R^2} \left(  \int_{0}^{\lambda_I} \frac{f_{\lambda}(\lambda,t)}{hc/\lambda}\    d\lambda \right) dt,
\end{equation}

\noindent
where 
$f_{\lambda}(\lambda,t)$ is the time series of observed spectral flux at Earth in erg\,s$^{-1}$\,cm$^{-2}$\,\AA$^{-1}$ (dereddened and deredshifted into the rest frame),
$\lambda_I$ is the wavelength of a photon at the ionization energy for species X in \AA,
$D$ is the distance to the SN from the Earth in cm,
$R$ is the distance to the material from the SN in cm (from the center of the SN, not the photosphere),
$t_0$ is the explosion time and $t$ is in seconds,
$\alpha_{\rm{X}}$ is the photoionization cross section for species X in cm$^{2}$,
and $N_{\rm{X}}(t)$ is the evolved column density of neutral species X at time $t$ in cm$^{-2}$ (i.e., $N_{\rm{X}}(t) = N_{\rm{X}}(t_0) - P_{\rm{X}}(t-dt)$).

We estimate the column density, $N_{\rm{K}}$, of \ion{K}{1} for the $v\approx-144$\,km\,s$^{-1}$ component for which pEW $\approx 30$\,m\AA\ by assuming this cloud is in the linear regime of the curve of growth, and find $N_{\rm{K}} \approx 8 \times 10^{10}$\,cm$^{-2}$. This agrees with Ritchey et al. (2014), who find $N_{\rm{K}}\approx5\times10^{10}$\,cm$^{-2}$ from a stack of their data (in which this component is not resolved and no evolution is detected), and that this same velocity component has $N_{\rm{Na}} \approx 28.5 \times 10^{11}$\,cm$^{-2}$. We use  approximate cross sections $\alpha_{\rm{K}} \approx 0.2 \times 10^{-18}$\,cm$^2$  and $\alpha_{\rm{Na}} \approx 0.1 \times 10^{-18}$\,cm$^2$ (e.g., Saha et al. 1988; Yeh \& Lindau 1985; Zatsarinny \& Tayal 2010). In lieu of an intrinsic NUV spectrum for SN\,2014J, and because Amanullah et al. (2014) have shown the intrinsic light curve of SN\,2014J to be very similar in width and peak brightness to that of SN\,2011fe, for $f_{\lambda}(\lambda,t)$ we use the time series of {\it HST} NUV spectra for SN\,2011fe from Mazzali et al. (2014) and $D=6.4$\,Mpc. 

\begin{figure}
\begin{center}
\includegraphics[width=8cm]{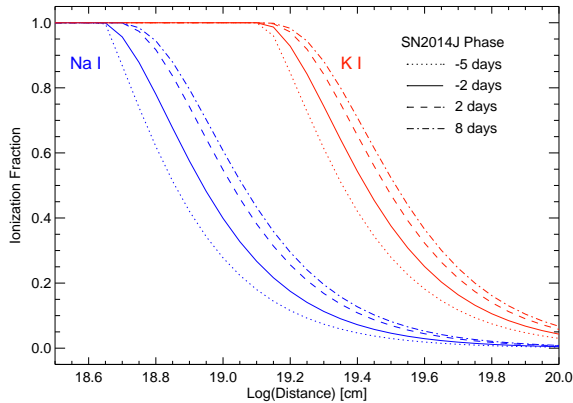}
\caption{A simple model of the cumulative fraction of \ion{Na}{1} (blue) and \ion{K}{1} (red) photoionized by SN\,Ia NUV radiation between explosion and the days around light-curve peak (solid, dash, dotted, and dot-dashed lines), assuming negligible recombination, as a function of distance between the SN\,Ia and the absorbing material. This shows that different fractions of the Na and K in nearby material can be photoionized by the SN\,Ia radiation.  \label{fig:ion}}
\end{center}
\end{figure}

In Figure \ref{fig:ion} we plot $P_{\rm{K}}$/$N_{\rm{K}}$ and $P_{\rm{Na}}$/$N_{\rm{Na}}$ around the time of light-curve peak, as a function of the distance between SN\,2014J and the absorbing material. Distances for which we see significant photoionization for \ion{K}{1} but not \ion{Na}{1} are $\sim$ (1--3) $\times 10^{19}$\,cm, or $\sim 3$--9\,pc. Unlike collisional ionization, this does not violate the limit of $\sim10^{17}$\,cm (Margutti et al. 2014; P\'{e}rez-Torres et al. 2014; Johansson et al. 2014), but it is not unambiguously close enough to be physically associated with the progenitor system and considered CSM (e.g., $<0.5$\,pc). At this distance, the material is not ionized by the shock breakout radiation, which has a significantly harder spectrum than the SN\,Ia (i.e., a blackbody with $T=2\times10^8$\,K) but a vastly lower total energy output of only $10^{40}$--$10^{44}$\,erg (e.g., H{\"o}flich \& Schaefer 2009; Piro et al. 2010). The rate of radiative recombination is long enough that we have neglected it is this calculation.

Although a smaller fraction of the Na would be photoionized in this scenario (e.g., $\sim20$\% at $\sim1.5 \times 10^{19}$\,cm), this would still be a significant amount of material --- a higher total number of ionized atoms than K. We do not see a $\sim$20\% reduction in the pEW of this velocity component of \ion{Na}{1}, which suggests that either our interpretation is incorrect or that this component of Na is not on the linear part of the curve of growth. Given the depth of the line and the saturation of many other components of Na, we consider the latter quite plausible. While our estimated distances are not quite low enough to declare this material circumstellar, we point out that only the scenario of CSM photoionization naturally explains why only the two most blueshifted components show evolution: the fastest-moving CSM is closest to the explosion, not yet decelerated by the ISM, and experiences a sufficiently strong radiation field. As a final note, we point out that there is precedence for a stronger effect of photoionization in \ion{K}{1}: Patat et al. (2011) show that during an outburst of the recurrent nova system RS~Oph --- an example of a potential progenitor system for SNe\,Ia --- the \ion{K}{1} absorption lines show a considerably faster decline in their equivalent width than the \ion{Na}{1} or \ion{Ca}{2}. 

\textbf{Collisional Ionization --} If this evolution in \ion{K}{1} stems from collisional ionization by high-velocity ejecta moving at $v\approx$ 15,000--20,000\,km\,s$^{-1}$ and reaching the CSM at $\sim10$--15 days after explosion, then the CSM material is $\sim 1.3$--2.6 $\times 10^{10}$\,km away from the explosion. This is closer than the closest material in the CSM of SN PTF\,11kx, which was at $\sim 1 \times 10^{11}$\,km, significantly closer than expected for swept-up ISM in the DD models ($\gtrsim3\times10^{12}$\,km; Shen et al. 2013), and in violation of the CSM limits from infrared and radio observations (e.g., $>10^{17}$\,cm away: Margutti et al. 2014; P\'{e}rez-Torres et al. 2014; Johansson et al. 2014). Most importantly, collisional ionization of the $v\approx-144$\,km\,s$^{-1}$ material would certainly be seen with equal strength in the \ion{Na}{1}~D component. For these reasons, we reject the possibility that we are seeing physical interaction between the ejecta and CSM.

\textbf{Transverse Motion --} Here we assume that all of the absorption features are caused by clouds of ISM. In this scenario, could SN\,2014J and an ISM cloud have a sufficiently large relative peculiar velocity to change the column density of K along our line of sight? Over the 30 days of our observations, relative peculiar velocities of 80--200\,km\,s$^{-1}$ could move our line of sight (2--5) $\times 10^8$\,km, or several AU, through a cloud. This would imply a very large density variation over a very small region, commonly referred to as tiny scale atomic structure (e.g., Heiles 1997). However, we would not expect a peculiar velocity effect for the most blueshifted (i.e., highest radial velocity) components; assuming the absorbing clouds are circularly rotating in the disk of M82, the most blueshifted clouds would be in the center of the disk and have the smallest relative peculiar velocities. But if the sightline to SN\,2014J has shifted out of this cloud, the corresponding component of \ion{Na}{1}~D should have completely dissipated also (unless large compositional variations are invoked), yet we do not observe this. We must therefore conclude that these line variations are not caused by the SN expanding behind a patchy ISM as described by Patat et al. (2010). Transverse motions are also ruled out for the shrinking pEW of the second-most blueshifted \ion{K}{1} feature at $v \approx -127$\,km\,s$^{-1}$ because it remains saturated in \ion{Na}{1}, indicating that the cloud covers the entire photosphere of the SN as seen from Earth for the duration of our observations.

\textbf{\ion{K}{1} Summary --} Despite potential telluric contamination and a bumpy continuum in the region of \ion{K}{1}, we tentatively conclude that the absorption features at $v\approx-144$ and $v\approx-127$\,km\,s$^{-1}$ show continuous, monotonic evolution. We find this observation difficult to attribute to an observational artifact, but also problematic to interpret as a physical phenomenon. We rule out transverse motions and collisional ionization, but find that a scenario in which the material is at a ``sweet spot," sufficiently near the SN to experience preferential photoionization of K over Na, is a somewhat contrived yet physically plausible explanation. Unfortunately, our data do not contain any lines that would help in this respect, such as resolved \ion{Ca}{1}, \ion{Ca}{2}, or \ion{K}{2}. Finally, we note that there have been no other reported observations of this phenomenon: neither Foley et al. (2014) nor Ritchey et al. (2014) describe evolution in the \ion{K}{1} 7665 complex (but our spectra have higher resolution and cover a broader range of phases).

\subsection{Calcium}

\begin{figure}
\begin{center}
\includegraphics[width=8cm]{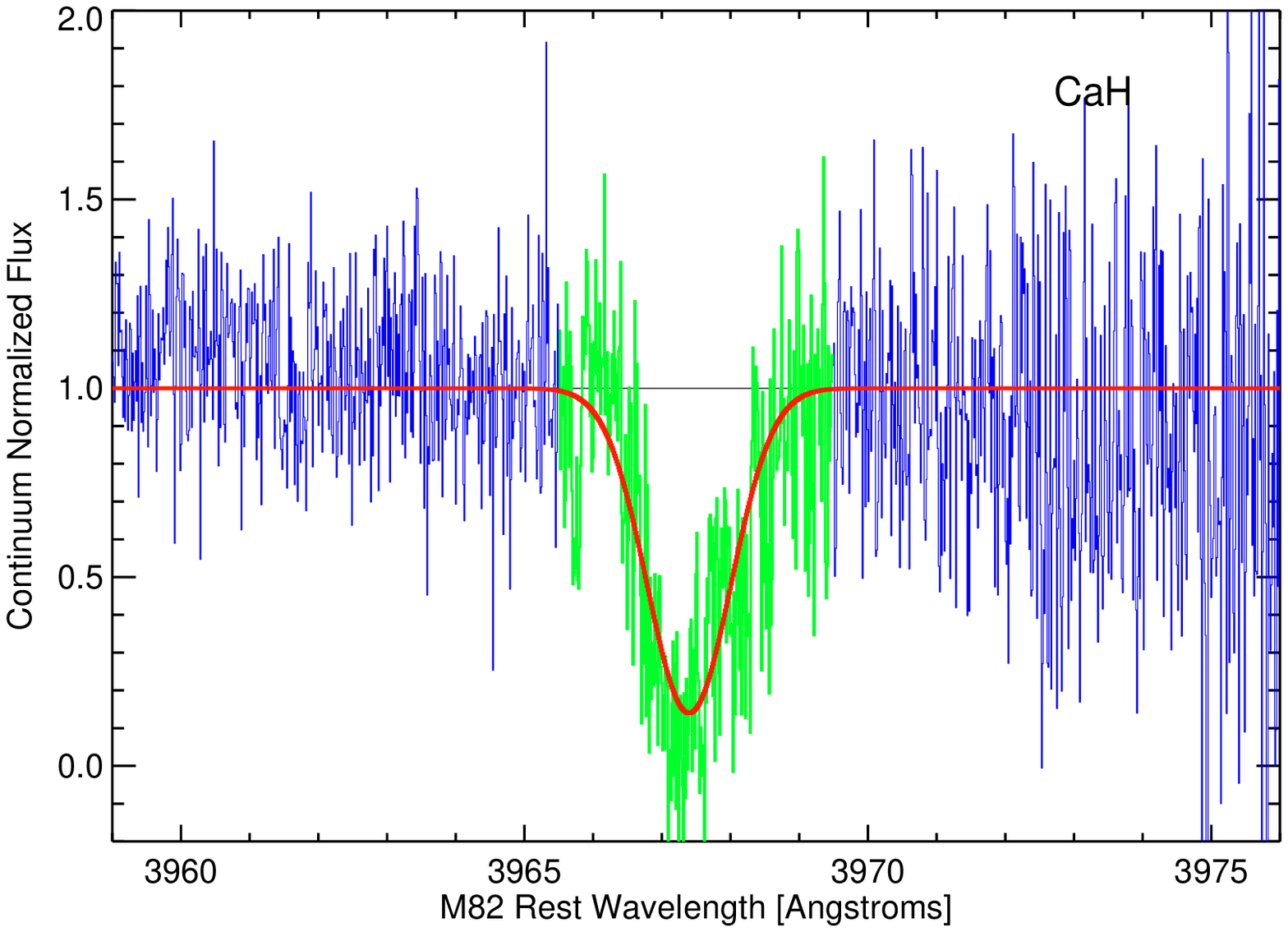}
\includegraphics[width=8cm]{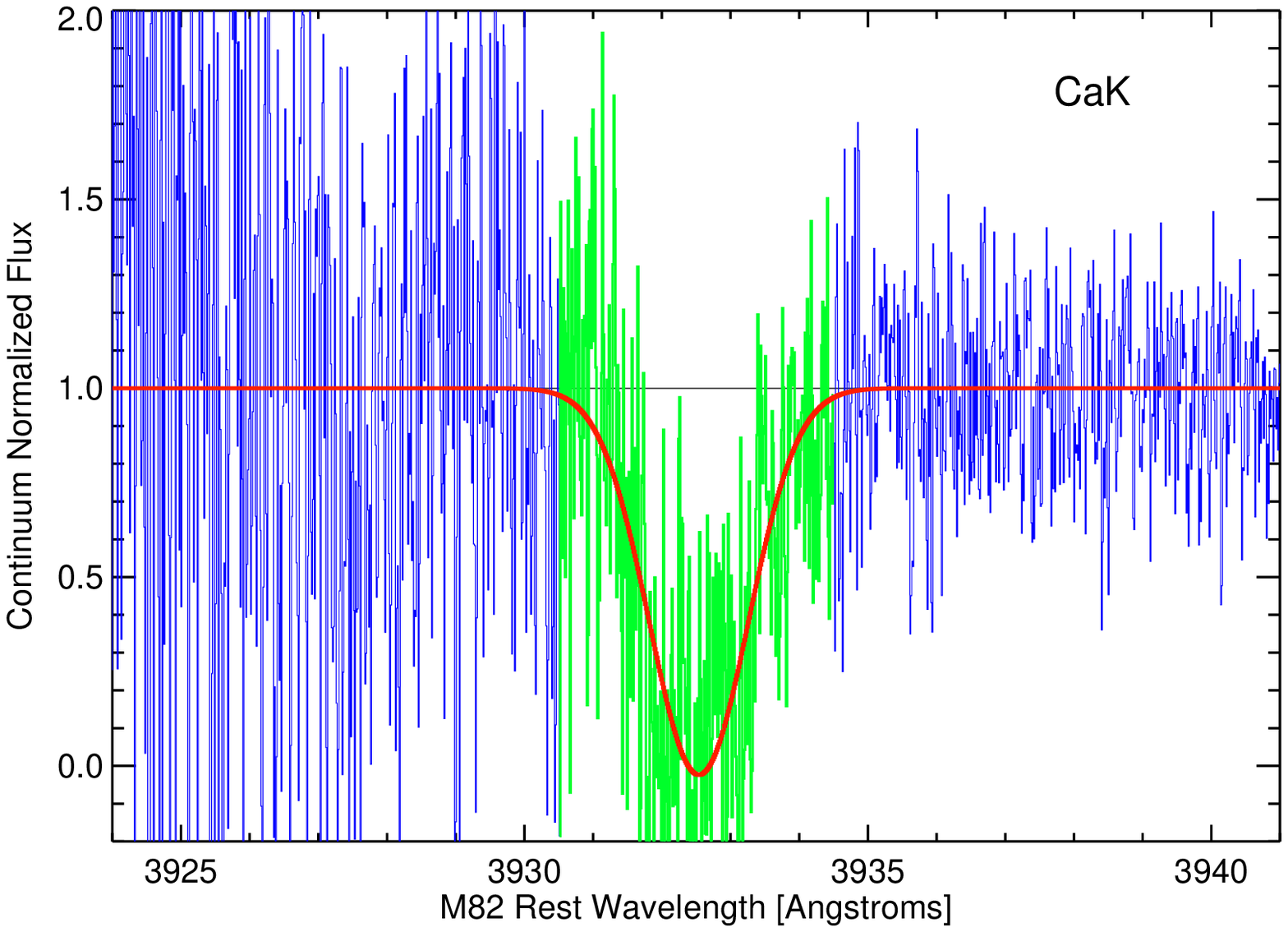}
\caption{Median-combined, continuum-normalized APF spectra for all epochs of SN\,2014J, showing the \ion{Ca}{2} H\&K lines. Blue represents the median spectrum, green represents the portion included in the fit, and the smooth red line is the best-fit Gaussian function. \label{fig:CaHK}}
\end{center}
\end{figure}

In Figure \ref{fig:CaHK} we show the median-combined, continuum-normalized spectra at the \ion{Ca}{2} H (3968.5\,\AA) and K (3933.7\,\AA) lines. We fit these lines with Gaussian profiles, the results of which are presented in Table \ref{tab:linepars}. We find that the equivalent widths are $1.31\pm0.25$\,\AA\ for \ion{Ca}{2} H, and $1.84\pm0.43$\,\AA\ for \ion{Ca}{2} K. Unfortunately, the sensitivity of the APF in the region of \ion{Ca}{2} H\&K is low, and we must stack the spectra to make any kind of robust measure of their equivalent width. This prohibits us from looking for evolution in the calcium features that might indicate CSM interaction, as was seen for PTF\,11kx (Dilday et al. 2012).

\begin{deluxetable}{lcc}
\tablecolumns{3}
\tablecaption{Gaussian Line Fit Parameters \label{tab:linepars}}
\tablehead{ \colhead{Source} & \colhead{FWHM} & \colhead{EW} \\ 
 & (\AA) & (\AA) }
\startdata
Ca H & 0.61 $\pm$ 0.07 & 1.31 $\pm$ 0.25                   \\ 
Ca K & 0.72 $\pm$ 0.09 & 1.84 $\pm$ 0.43                   \\ 
DIB 5780 & 1.19 $\pm$ 0.05 & 0.34 $\pm$ 0.02  \\ 
DIB 5797 & 0.78 $\pm$ 0.04 & 0.21 $\pm$ 0.02  \\ 
DIB 6196 & 0.50 $\pm$ 0.07 & 0.05 $\pm$ 0.01  \\ 
DIB 6283 & 1.47 $\pm$ 0.05 & 0.36 $\pm$ 0.02  \\ 
DIB 6613 & 1.12 $\pm$ 0.06 & 0.26 $\pm$ 0.02
\enddata
\tablenotetext{}{}
\end{deluxetable}

\subsection{Diffuse Interstellar Bands}

Merrill et al. (1934) first identified interstellar spectral absorption lines that, unlike the narrow lines of \ion{Na}{1}~D and \ion{Ca}{2} H\&K, were broad and had nonsharp edges. The sources of these lines remain unidentified, and they are commonly referred to as diffuse interstellar bands (DIBs). In the past, the established relation between the EW of the \ion{Na}{1}~D absorption feature and $A_V$ has typically been used to estimate the extinction of a SN\,Ia. However, Phillips et al. (2013) find that 25\% of SNe\,Ia have stronger \ion{Na}{1}~D than expected from their extinction ($A_V$), and they show that the EW of the DIB at 5780\,\AA\ is more correlated with SN\,Ia dust extinction:
\begin{equation}
\log{{\rm EW}_{5780}} = 2.283 + \log{A_V}.
\end{equation}
\noindent
This relation has ``a 50\% error in $A_V$ if the 5780\,\AA\ feature is used to estimate the dust extinction for any single object" (Equation 6 of Phillips et al. 2013), and is particularly useful as an independent estimate of $A_V$ in cases where the \ion{Na}{1}~D line is saturated. 

\begin{figure}
\begin{center}
\includegraphics[trim=0.5cm 0.3cm 0.6cm 0.3cm,clip=true,width=5.77cm]{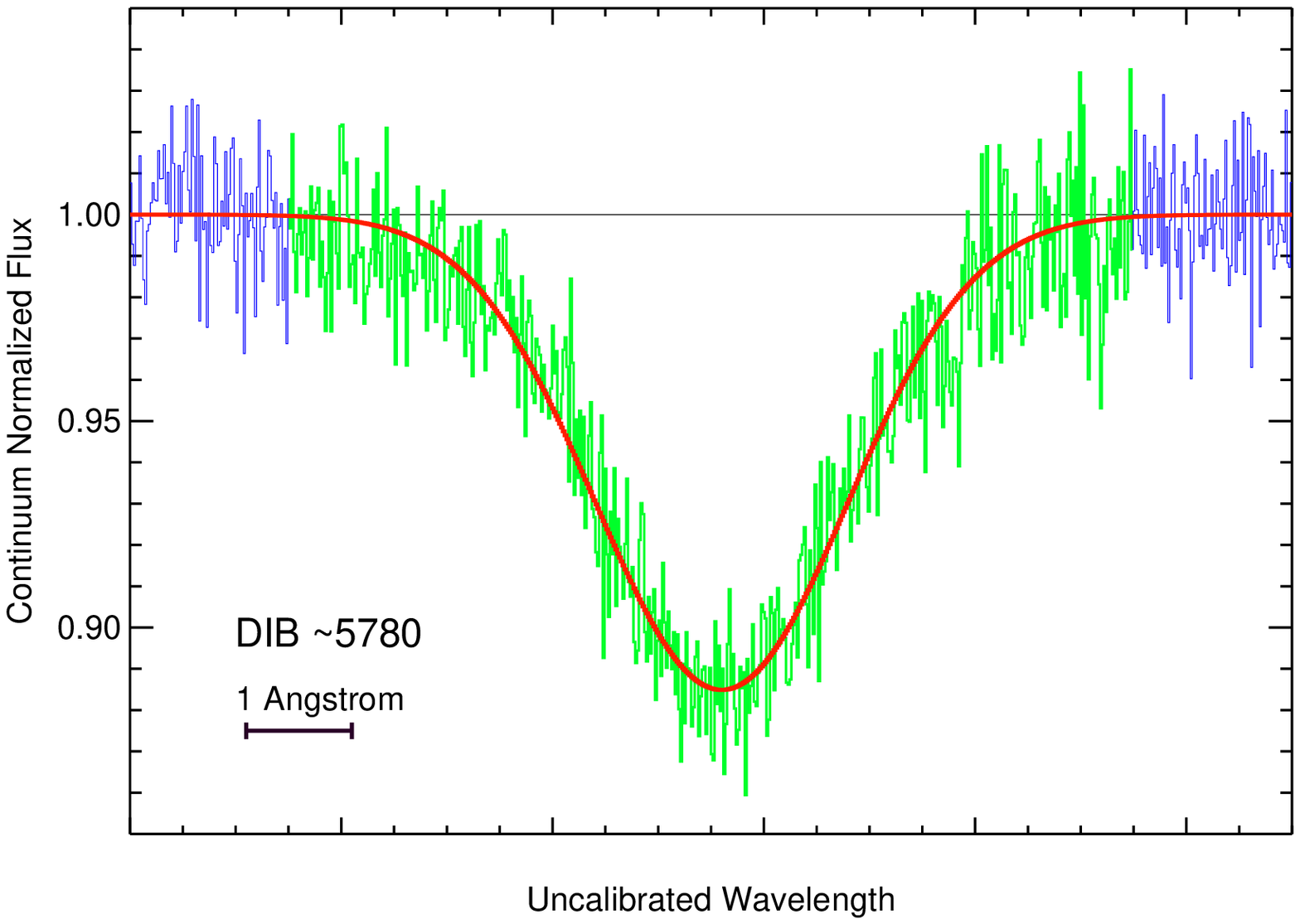}
\includegraphics[trim=0.5cm 0.3cm 0.6cm 0.3cm,clip=true,width=5.77cm]{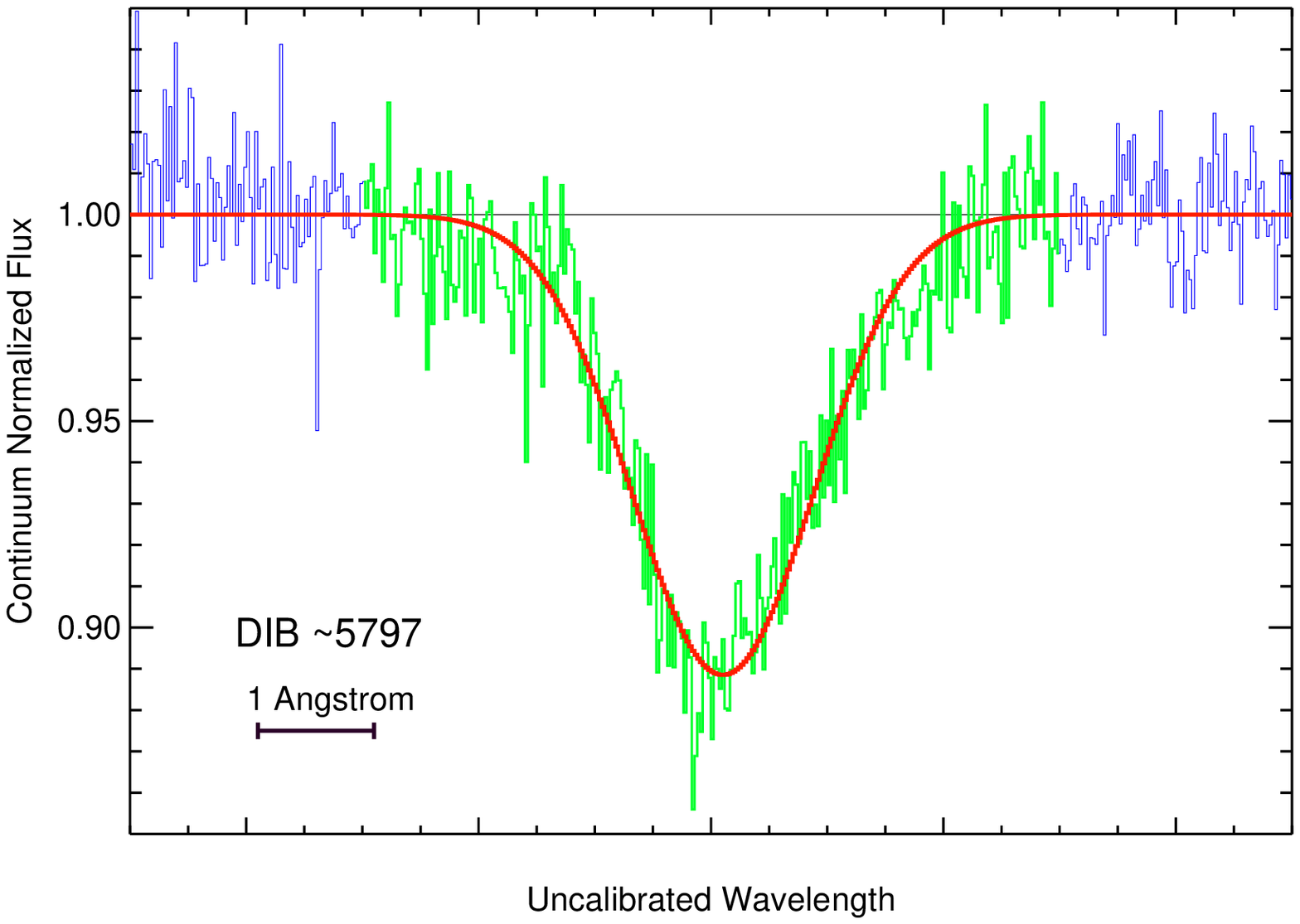}
\includegraphics[trim=0.5cm 0.3cm 0.6cm 0.3cm,clip=true,width=5.77cm]{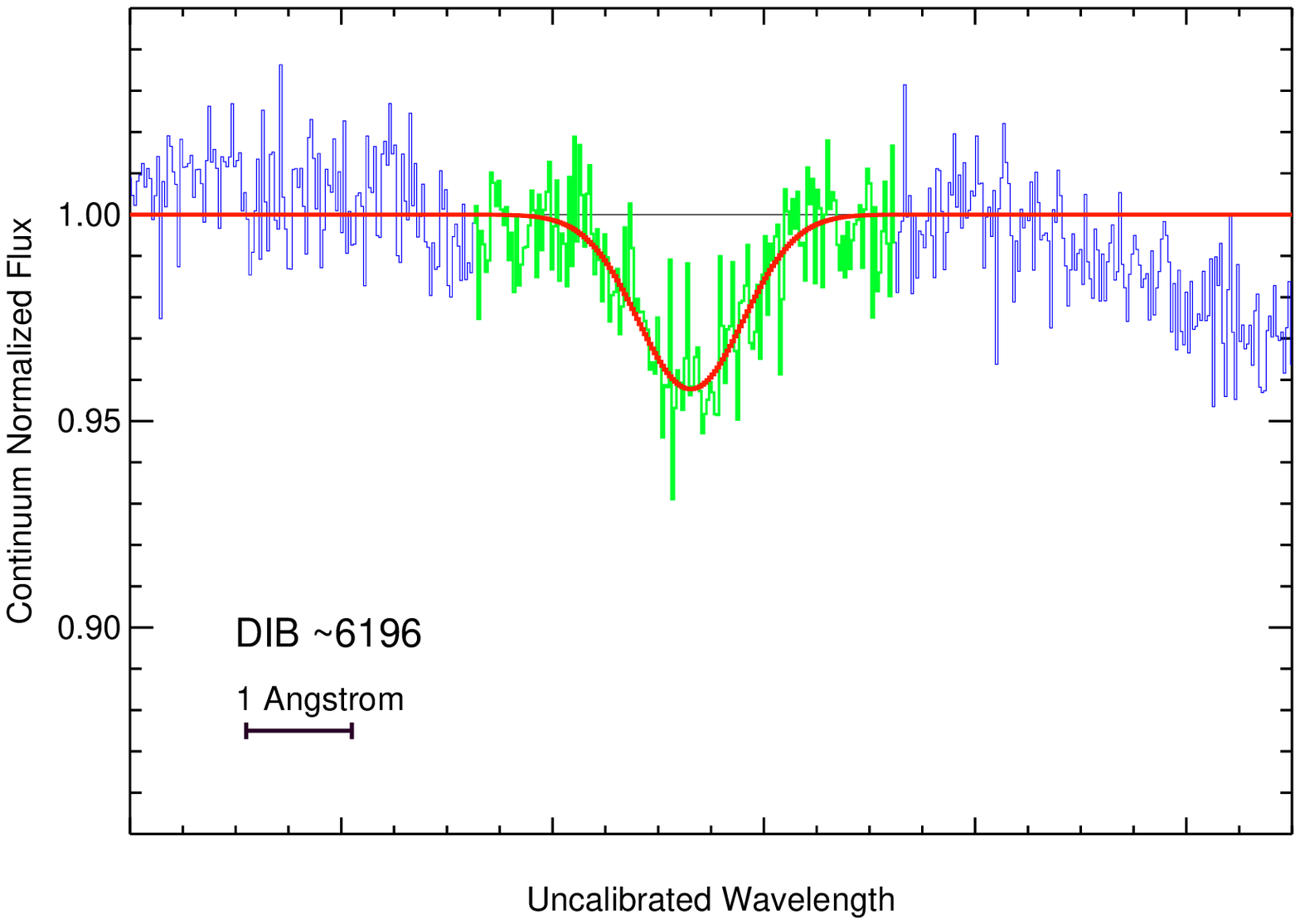}
\includegraphics[trim=0.5cm 0.3cm 0.6cm 0.3cm,clip=true,width=5.77cm]{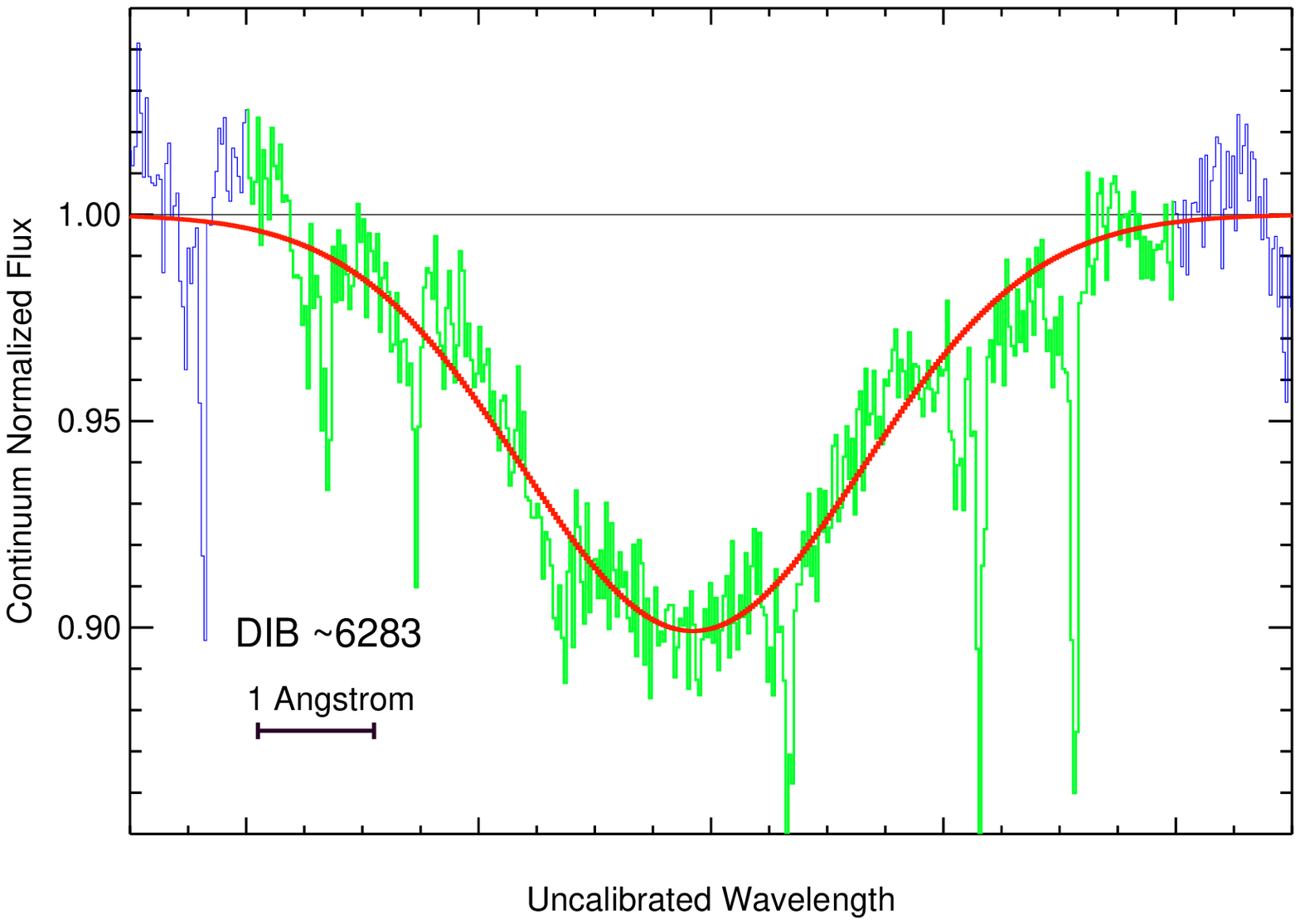}
\includegraphics[trim=0.5cm 0.3cm 0.6cm 0.3cm,clip=true,width=5.77cm]{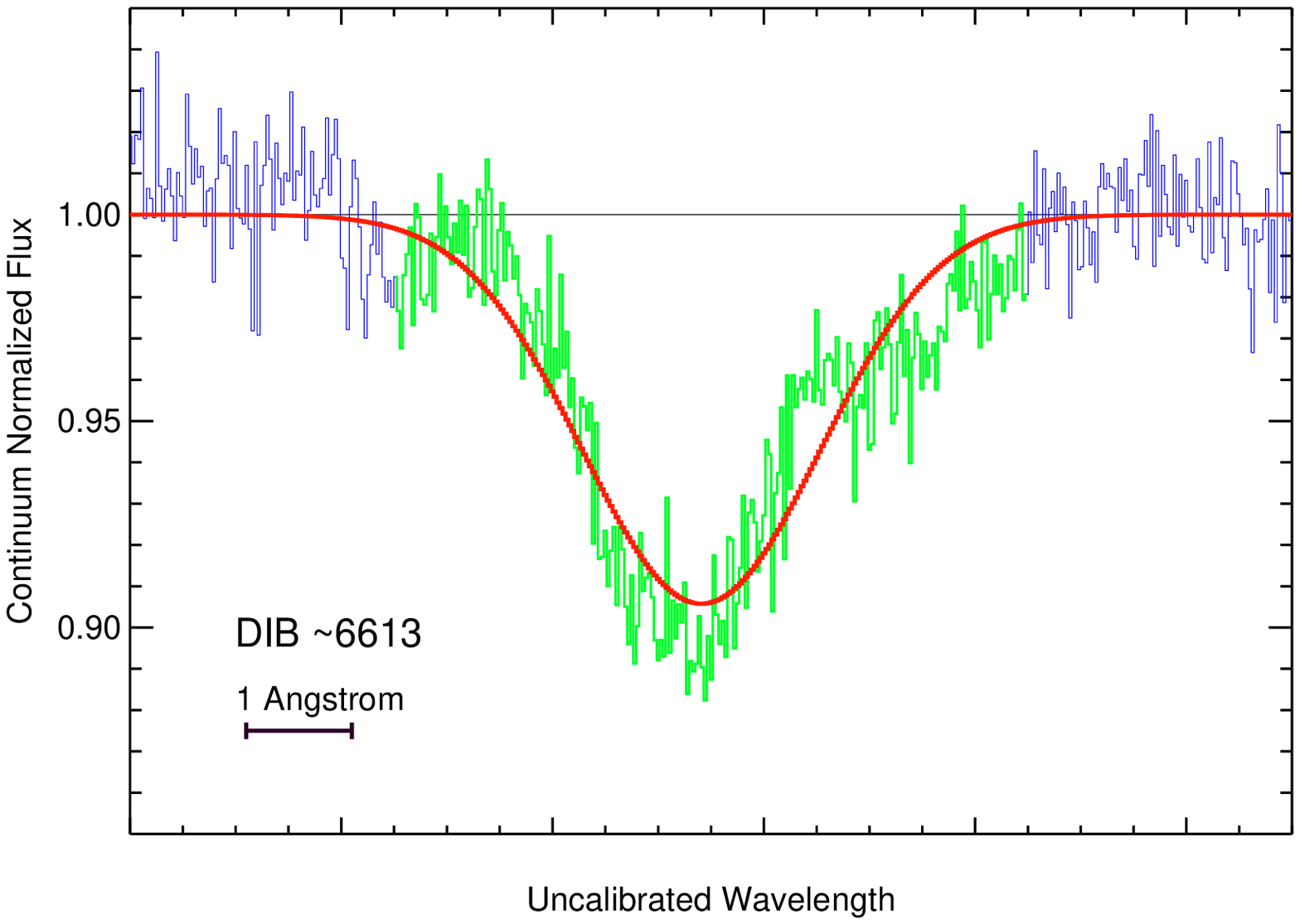}
\caption{Median-combined, continuum-normalized APF spectra for all epochs of SN\,2014J at the positions of the five DIBs we detect. Blue represents the median spectrum, green represents the portion included in the fit, and the smooth red line is the best-fit Gaussian function. The ordinate axes of the plots are matched in scale to show the relative DIB strengths.  \label{fig:DIB}}
\end{center}
\end{figure}

In Figure \ref{fig:DIB} we show the continuum-normalized, median-combined spectra of the five DIBs that we detect in the spectrum of SN\,2014J. As with the calcium lines, we have fit the median spectrum with a Gaussian function, and the fit parameters are listed in Table \ref{tab:linepars}. We find that EW$_{\lambda5780} = 0.34 \pm 0.02$\,\AA\ (Table \ref{tab:linepars}). Through Equation 1, this implies $A_V^{\rm{host}} = 1.8\pm0.9$\,mag. This agrees well with the value determined by Foley et al. (2014), but disagrees with the values presented by Goobar et al. (2014) of EW$_{\lambda5780} = 0.48 \pm 0.01$\,\AA\ and $A_V^{\rm{host}} = 2.5\pm1.3$\,mag. 

\begin{figure}[h]
\begin{center}
\epsscale{1.2}
\plotone{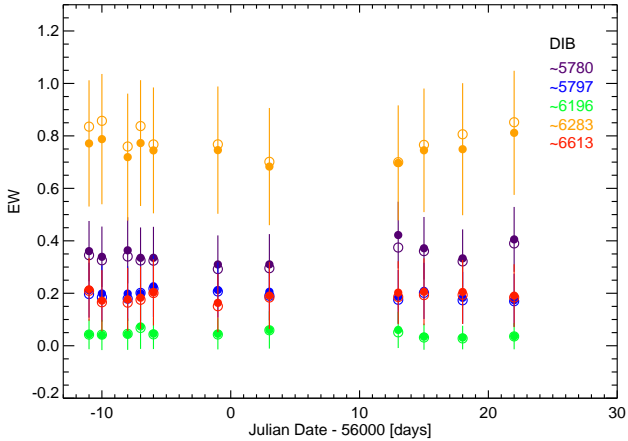}
\caption{The equivalent width (EW) of each identified DIB as a function of time. Filled circles are the EW from the Gaussian fits to spectra (shown with their error bars), and open circles are from direct integrations of the data (shown without their  error bars). None of the DIBs shows any significant evolution in EW. \label{fig:dibew}}
\end{center}
\end{figure}

Recently, a time series of high-resolution spectra of SN\,Ic-BL 2012ap revealed exceptionally strong DIBs that evolved over 30 days, suggesting that the DIB carriers have a high ionization potential, that their emission is sensitive to the SN radiation, and that they may be cations or charged fullerenes \citep{milisav2014}. To look for evolution in the DIBs, in Figure \ref{fig:dibew} we plot the EW as a function of time, and find it consistent with no evolution in the DIBs. This agrees with Figure 4 of Foley et al. (2014), who show the EW of DIBs at 5780 and 5797\,\AA\ for phases $-10$ to +40 days and find no change.

\section{Conclusions} \label{sec:disc}

We present a time series of spectra with the highest resolution yet for SN\,Ia 2014J. We find multiple absorbing components of \ion{Na}{1}~D and \ion{K}{1}, as well as absorption by \ion{Ca}{2} H\&K and several DIBs. For \ion{Na}{1} and \ion{K}{1}, we see that most of these components are blueshifted with velocities up to $-150$\,km\,s$^{-1}$ relative to the recession velocity of M82, which is consistent with the location of SN\,2014J on the approaching side of the disk. 

Surprisingly, we find that the two most blueshifted absorption features of \ion{K}{1} weaken/dissipate during our spectral time series, but the corresponding velocity components of \ion{Na}{1}~D do not. We find that collisional ionization and transverse motion are less likely to be the cause of the variation in \ion{K}{1}, and instead show that photoionization by NUV light from SN\,2014J is a plausible explanation if the material is $\sim10^{19}$\,cm from the SN. The fact that only the most blueshifted features exhibit evolution is consistent with a physical model in which CSM is released by the SN\,Ia progenitor system during the late stages of its evolution. The most recently released material would have the highest velocity (not yet decelerated by the ISM), and be closest to the progenitor and more likely to experience photoionization. However, our distance estimates are slightly too large to confirm that the material exhibiting the evolving \ion{K}{1} is truly circumstellar in nature.

Under this interpretation, can we place any constraints on the progenitor scenario for SN\,2014J? Blueshifted material is \textit{not} a unique identifier for the SD scenario, as it is also a natural byproduct of the later stages of DD binary evolution when the companion is a He WD (Shen et al. 2013). Despite this, \textit{multiple} components of CSM are not (yet) a prediction from any DD scenario model; consequently, we find that our results tentatively support a SD scenario for SN\,2014J. However, as with most other SNe\,Ia, we ultimately concede that without a direct detection of hydrogen there is no clear evidence for a nondegenerate companion.

\acknowledgments

{\it Facilities:} \facility{Lick Observatory: Automated Planet Finder Telescope}.

Based on observations made with the Levy Spectrometer on the 2.4\,m Automated Planet Finder telescope at Lick Observatory. We thank Ken and Gloria Levy for their generous contribution to the funding of the spectrometer, without which this unique time series of spectra of SN\,2014J in M82 would have been impossible. We gratefully acknowledge Ferdinando Patat, Helen Kirk, Sarah Sadavoy, Christopher F. McKee, and Richard McCray for helpful discussions. The supernova research of A.V.F.'s group at U.C. Berkeley is supported by Gary \& Cynthia Bengier, the Richard \& Rhoda Goldman Fund, the Christopher R. Redlich Fund, the TABASGO Foundation, and NSF grant AST--1211916. We also acknowledge NASA/HST grant GO--13286 from the Space Telescope Science Institute, which is operated by the Association of Universities for Research in Astronomy, Inc., under NASA contract NAS 5--26555. K.J.S. is supported by NASA through the Astrophysics Theory Program (NNX15AB16G). This work has also used the Weizmann Interactive Supernova Data Repository (WISEREP) at www.weizmann.ac.il/astrophysics/wiserep.

\bibliographystyle{apj}
\bibliography{apj-jour,mybib}


\begin{thebibliography}{10}
\expandafter\ifx\csname natexlab\endcsname\relax\def\natexlab#1{#1}\fi

\bibitem[{Amanullah et al.}(2014)]{Aman2014}
Amanullah, R. 2014, \apj, 788, 21. 

\bibitem[{Blondin et al.}(2009)]{Blondin2009}
{Blondin}, S., {Prieto}, J.~L., {Patat}, F., {Challis}, P.,	{Hicken}, M., {Kirshner}, R.~P., {Matheson}, T. and {Modjaz}, M. 2009, \apj, 693, 207.

\bibitem[{Cao et al.}(2014)]{Cao2014}
{Cao}, Y., {Kasliwal}, M. M., {McKay}, A. and {Bradley}, A. 2014, ATel, 5786, 1.

\bibitem[{Cox \& Patat}(2008)]{Cox2008}
{Cox}, N. \& {Patat}, N. 2008, \aap, 485, 9.

\bibitem[{de Vaucouleurs et al.}(1991)]{devac1991}
{de Vaucouleurs}, G., {de Vaucouleurs}, A., {Corwin}, Jr., H.~G., {Buta}, R.~J., {Paturel}, G. and {Fouqu{\'e}}, P. 1991, Third Reference Catalogue of Bright Galaxies, V9.

\bibitem[{Dilday et al.}(2012)]{Dilday2012}
Dilday, B. et al. 2012, Science, 337, 942.

\bibitem[{D'Odorico et al.}(1989)]{DOdorico1989}
{D'Odorico}, S., {di Serego Alighieri}, S., {Pettini}, M., {Magain}, P., {Nissen}, P.~E. and {Panagia}, N. 1989, \aap, 215, 21.

\bibitem[{Fossey et al.}(2014)]{Fossey2014}
Fossey, S.~J., et al. 2014, CBET, 3792

\bibitem[{Foley et al.}(2014)]{Foley2014}
Foley, R.~J. et al. 2014, \mnras, 443, 2887.

\bibitem[{Goobar et al.}(2014)]{Goobar2014}
Goobar, A. et al. 2014, \apj, 784, 12.

\bibitem[{Goobar et al.}(2014b)]{Goobar2014b}
Goobar, A. et al. 2014b, [arXiv:1410.1363]

\bibitem[{Heiles}(1997)]{Heiles1997}
{Heiles}, C. 1997, \apj, 481, 193.

\bibitem[{H{\"o}flich \& Schaefer}(2009)]{Hoeflich2009}
{H{\"o}flich}, P. and {Schaefer}, B.~E., 2009, \apj, 705, 483.

\bibitem[{Howell}(2011)]{Howell2011}
{Howell}, D.~A. 2011, Nature Communications, 2, 350.

\bibitem[{Jacobs et al.}(2009)]{Jacobs2009}
{Jacobs}, B.~A., {Rizzi}, L., {Tully}, R.~B., {Shaya}, E.~J., {Makarov}, D.~I. and {Makarova}, L. 2009, \aj, 138, 332.

\bibitem[{Johansson et al.}(2014)]{Johansson2014}
{Johansson}, J. et al. 2014, \mnras, submitted, [arXiv:1411.3332]

\bibitem[{Kawabata et al.}(2014)]{Kawabata2014}
Kawabata, K.~S. et al. 2014, \apj, 795, 4.

\bibitem[{Kelly et al.}(2014)]{Kelly2014}
Kelly, P. et al. 2014, \apj, 790, 3.

\bibitem[{Kramida et al.}(2013)]{NIST}
Kramida, A., Ralchenko, Yu., Reader, J., and NIST ASD Team (2013). NIST Atomic Spectra Database (ver. 5.1), National Institute of Standards and Technology, Gaithersburg, MD. 



\bibitem[{Maguire et al.}(2013)]{Maguire2013}
Maguire et al. 2013, \mnras, 436, 222.

\bibitem[{Mandel et al.}(2014)]{Mandel2014}
{Mandel}, K.~S., {Foley}, R.~J. and {Kirshner}, B.~P. 2014, \apj, submitted [arXiv:1402.7079]

\bibitem[{Margutti et al.}(2014)]{Margutti2014}
Margutti, R., Parrent, J., Kamble, A., Soderberg, A. M., Foley, R. J., Milisavljevic, D., Drout, M. R. and Kirshner, R. 2014, \apj, 790, 52.

\bibitem[{Mauerhan}{\ et al.\ }(2013)]{Mauerhan2013}
{Mauerhan}, J.~C. 2013, \mnras, 430, 1801.

\bibitem[{Mazzali et~al}(2014)]{Mazzali2014}
{Mazzali}, P.~A. et al. 2014, \mnras, 439, 1959

\bibitem[{Merrill}(1934)]{Merrill1934}
{Merrill}, P.~W. 1934, \pasp, 46, 206

\bibitem[{Milisavljevic et al.}(2014)]{milisav2014}
{Milisavljevic}, D. et al. 2014, \apj L, 782, 5.

\bibitem[{Milne et al.}(2013)]{Milne2013}
{Milne} P.~A., {Brown} P.~J., {Roming} P.~W.~A., {Bufano} F., {Gehrels} N., 2013, \apj, 779, 23.

\bibitem[{Nielsen et al.}(2014)]{Nielsen2014}
{Nielsen}, M.~T.~B., {Gilfanov}, M., {Bogd\'{a}n}, \'{A}., {Woods}, T.~E., {Nelemans}, G. 2014, \mnras, 442, 3400.

\bibitem[{Patat et al.}(2007)]{Patat2007}
Patat, F. et al. 2007, Science, 317, 924.

\bibitem[{Patat et al.}(2010)]{Patat2010}
{Patat}, F., {Cox}, N.~L.~J., {Parrent}, J. and {Branch}, D. 2010, \aap, 514, 78.

\bibitem[{Patat et al.}(2011)]{Patat2011}
{Patat}, F., {Chugai}, N.~N., {Podsiadlowski}, P., {Mason}, E., {Melo}, C. and {Pasquini}, L. 2011, \aap, 530, 63.

\bibitem[{Patat et al.}(2013)]{Patat2013}
Patat, F. et al. 2013, \aap, 549, 62.

\bibitem[{Patat et al.}(2014)]{Patat2014}
{Patat}, F. et al. 2014, \aap, submitted, [arXiv:1407.0136]

\bibitem[{P\'{e}rez-Torres et al.}(2014)]{PT2014}
{P\'{e}rez-Torres}, M.~A. et al. 2014, \apj, 792, 38. 

\bibitem[{Pettini}(1988)]{Pettini1988}
{Pettini}, M. 1988, PASAu, 7, 527.

\bibitem[{Phillips et al.}(2013)]{Phillips2013}
{Phillips}, M. et al. 2013, \apj, 779, 38.

\bibitem[{Piro et al.}(2010)]{Piro2010}
{Piro}, A.~L., {Chang}, P. and {Weinberg}, N.~N. 2010, \apj, 708, 598.

\bibitem[{Rich}(1987)]{Rich1987}
{Rich} R.~M. 1987, \aj, 94, 3.

\bibitem[{Ritchey et al.}(2014)]{Ritchey2014}
{Ritchey}, A.~M., {Welty}, D.~E., {Dahlstrom}, J.~A. and {York}, D.~G. 2014, \apj, submitted [arXiv:1407.5723]

\bibitem[{Saha et al.}(1988)]{Saha1988}
{Saha} H.~P., {Froese-Fischer} C. \& {Langhoff} P.~W. 1988, \pra, 38, 1279.

\bibitem[{Shen et al.}(2013)]{Shen2013}
{Shen}, K.~J., {Guillochon}, J., \& {Foley}, R.~J. 2013, \apj L, 770, 35.

\bibitem[{Simon et al.}(2009)]{Simon2009}
{Simon}, J.~D. et al. 2009, \apj, 702, 1157.

\bibitem[{Sollerman et al.}(2005)]{Sollerman2005}
{Sollerman} J., {Cox}, N., {Mattila}, S., {Ehrenfreund}, P., {Kaper}, L., {Leibundgut}, B. and {Lundqvist}, P. 2005, \aap, 429, 559.

\bibitem[{Steidel et al.}(1990)]{Steidel1990}
{Steidel}, C.~C., {Rich}, R.~M. and {McCarthy}, J.~K. 1990 \aj, 99, 5.

\bibitem[{Sternberg et al.}(2011)]{Sternberg2011}
{Sternberg}, A. et al. 2011, Science, 333, 856.

\bibitem[{Sternberg et al.}(2014)]{Sternberg2014}
{Sternberg}, A. et al. 2014, \mnras, 443, 1849.


\bibitem[{Vladilo et al.}(1987a)]{Vladilo1987a}
{Vladilo} et al. 1987a, \aap, 177, 17 

\bibitem[{Vladilo et al.}(1987b)]{Vladilo1987b}
{Vladilo} et al. 1987b, \aap, 182, 59 

\bibitem[{Vogt et al.}(2014)]{Vogt2014}
{Vogt}, S.~S. et al. 2014, \pasp, 126, 359.

\bibitem[Wang (2005)]{Wang2005}
{Wang} L. et al. 2005, \apj, 635, 33.

\bibitem[{Wang et al.}(2013)]{Wang2013}
{Wang} L. et al. 2013, Science, 430, 170.

\bibitem[{Welty et al.}(2014)]{Welty2014}
{Welty}, D.~E., {Ritchey}, A.~M., {Dahlstrom}, J.~A. and {York}, D.~G. 2014, \apj, 792, 106.

\bibitem[{Westmoquette et al.}(2009)]{Westm2009}
{Westmoquette}, M.~S., {Smith}, L.~J., {Gallagher}, J.~S., {Trancho}, G., {Bastian}, N. and {Konstantopoulos}, I.~S. 2009, \apj, 696, 192

\bibitem[{Westmoquette et al.}(2013)]{Westm2013}
{Westmoquette}, M.~S., {Smith}, L.~J., {Gallagher}, J.~S. and {Walter} F. 2013, MNRAS 428, 1743.

\bibitem[{Yaron} \& {Gal-Yam}(2012)]{2012PASP..124..668Y}
{Yaron} O., {Gal-Yam} A., 2012, \pasp, 124, 668.

\bibitem[{Yeh \& Lindau}(1985)]{Yeh1985}
{Yeh}, J.~J. \& {Lindau}, I. 1985, Atomic Data and Nuclear Data Tables, 32, 1.

\bibitem[{Zatsarinny \& Tayal}(2010)]{Zats2010}
{Zatsarinny} O. \& {Tayal} S.~S. 2010, \pra, 81, 043423.

\bibitem[{Zheng et al.}(2014)]{Zheng2014}
Zheng, W. et al. 2014, \apj L, 783, 24.




\end{thebibliography}

\end{document}